\documentclass[a4paper]{jpconf}

\usepackage{graphics}
\usepackage{graphicx}
\usepackage{bm}
\usepackage{amsmath}
\usepackage{amsfonts} 
\usepackage{amssymb}
\usepackage{latexsym}
\usepackage{color}
\usepackage{mathrsfs}
\usepackage{braket}
\usepackage{enumerate}
\usepackage[usenames,dvipsnames]{xcolor}
\usepackage[colorlinks=true,citecolor=blue,linkcolor=blue,urlcolor=blue,backref=true]{hyperref}
\usepackage{float}
\usepackage{mathtools}
\usepackage[raggedright,tight]{subfigure}  %side-by-side figures
\begin{document}

\title{A master equation for strongly interacting dipoles}
\author{Adam Stokes$^1$}
\author{Ahsan Nazir$^1$}

\address{$^{1}$Photon Science Institute and School of Physics \& Astronomy, The University of Manchester, Oxford Road, Manchester, M13 9PL, United Kingdom}

\begin{abstract}
We consider a pair of dipoles such as Rydberg atoms for which direct electrostatic dipole-dipole interactions may be significantly larger than the coupling to transverse radiation.  We derive a master equation using the Coulomb gauge, which naturally enables us to include the inter-dipole Coulomb energy within the system Hamiltonian rather than the interaction. In contrast, the standard master equation for a two-dipole system, which depends entirely on well-known gauge-invariant $S$-matrix elements, is usually derived using the multipolar gauge, wherein there is no explicit inter-dipole Coulomb interaction. We show using a generalised arbitrary-gauge light-matter Hamiltonian that this master equation is obtained in other gauges only if the inter-dipole Coulomb interaction is kept within the interaction Hamiltonian rather than the unperturbed part as in our derivation. Thus, our master equation depends on different $S$-matrix elements, which give separation-dependent corrections to the standard matrix elements describing resonant energy transfer and collective decay. The two master equations coincide in the large separation limit where static couplings are negligible. We provide an application of our master equation by finding separation-dependent corrections to the natural emission spectrum of the two-dipole system.
\end{abstract}
 
\section{Introduction}\label{int}

Dipole-dipole interactions are central to several important effects in atomic and molecular physics. Early studies by Eisenschitz, London and F\"orster \cite{london_zur_1930,forster_zwischenmolekulare_1948} treated dipolar interactions as perturbative effects arising from direct electrostatic coupling. Molecular quantum electrodynamics (QED) extends these treatments by incorperating retardation effects due to finite signal propagation.  As was first shown by Casimir and Polder \cite{casimir_influence_1948}, a striking retardation effect occurs at large separations $R/\lambda \gg 1$ where the $R^{-6}$ dependence of the dispersion energy is increasingly replaced by an $R^{-7}$ dependence.

In order to study the dynamics of systems of interacting dipoles open quantum systems theory has proven useful \cite{agarwal_quantum_2012}. 
The master equation formalism can be used to obtain dynamical information about state populations and coherences, and to obtain fluorescence spectra \cite{freedhoff_collective_1979,kilin_cooperative_1980,griffin_two-atom_1982,ficek_two-atom_1990}. As will be confirmed in this work, the standard second-order Born-Markov-secular master equation describing two dipoles within a common radiation reservoir depends entirely on well-known quantum electrodynamic (QED) matrix elements. These matrix elements describe dipole-dipole coupling and decay with retardation effects included. This master equation is obtained by treating the direct electrostatic coupling between the dipoles as a perturbation along with the coupling to transverse radiation. However, it is clear that if the former is sufficiently strong this approach may not be justified, in analogy with the case of externally imposed interactions~\cite{santos_master_2014}. Here we consider a system of free dipoles strongly coupled by dipole-dipole interactions. Our focus is on discerning the full dependence of the physics on the inter-dipole separation. We also delineate how microscopic gauge-freedom effects the ensuing master equation derivation.

An important class of systems strongly coupled by dipole-dipole interactions are Rydberg atoms, which have been of interest for some time \cite{raimond_spectral_1981}. In recent years dipole-dipole interactions of Ryberg atoms have been the subject of numerous experimental and theoretical works \cite{reinhard_level_2007,vogt_electric-field_2007,altiere_dipole-dipole_2011,ravets_coherent_2014,zhelyazkova_probing_2015,bachor_addressing_2016,kumar_collective_2016,dyachkov_dipoledipole_2016,browaeys_experimental_2016,petrosyan_grover_2016,saffman_quantum_2016,dunning_recent_2016,paris-mandoki_tailoring_2016}. Recently the first experimental confirmation of F\"orster resonant energy transfer was demonstrated using two Rydberg atoms separated by $15\mu m$ \cite{ravets_coherent_2014}. This type of resonant energy transfer is an important mechanism within photosynthesis, whose quantum nature is of continued interest within open quantum systems theory \cite{ishizaki_quantum_2010}. Dipole-dipole interactions of Rydberg atoms also offer promising means of implementing quantum gates in which adjacent Rydberg states are treated as effective two-level systems and dipolar interactions are tuned with the use of lasers \cite{saffman_quantum_2016}.

Such adjacent Rydberg states are typically separated by microwave transitions, which for small enough separations can be matched or even exceeded by the electrostatic dipole-dipole interaction strength divided by $\hbar$. Thus, a novel regime of {\em strong electrostatic coupling} occurs, in which the usual weak-coupling theory is expected to break down. A repartitioning of the Hamiltonian is necessary in order to identify a genuinely weak system-reservoir interaction, which can then constitute the starting point for perturbation theory. More specifically, we include the direct inter-dipole Coulomb energy within the unperturbed part of the Hamiltonian and only treat the coupling to transverse radiation as a weak perturbation. The master equation we derive exhibits a different dependence on the inter-dipole separation, and this has important consequences for the predicted physics. The rates of collective decay and resonant energy transfer are altered, as are the properties of the light emitted by the system. 

There are five sections in this paper. We begin in Section \ref{1} by reviewing the standard one and two dipole master equations in the Born-Markov and secular approximations. We show how the standard two-dipole master equation can be obtained for various choices of gauge for the microscopic Hamiltonian. Our purpose is to clearly identify limitations in the standard derivation, which is usually always performed using the multipolar Hamiltonian \cite{agarwal_quantum_2012}. This concrete form of the Hamiltonian is the form obtained by choosing the multipolar gauge, also known as the Poincar\'e gauge \cite{cohen-tannoudji_photons_1989}. In Section \ref{2} we derive an alternative master equation describing the two-dipole system, which only reduces to the standard result in the limit of vanishing direct electrostatic coupling between the dipoles. This occurs in the limit of large separation. In Section \ref{3} we solve the master equation derived in Section \ref{2} and compare the solution with that of the standard master equation. We also obtain corrections to the emission spectrum of the two-dipole system. Finally in Section \ref{5} we summarise our findings. We assume natural units $\hbar=\epsilon_0=c=1$ throughout.

\section{Gauge-invariant master equations}\label{1}

\subsection{Single-dipole Hamiltonian and master equation}\label{1.1}

Here we identify sufficient conditions in order that the same master equation can be obtained from different microscopic Hamiltonians. This will be important when it comes to deriving the two-dipole master equation in the following sections. Let us consider a single dipole within the electromagnetic bath, and assume that there are only two relevant states ($\ket{g}, \ket{e}$) of the dipole separated by energy $\omega_0=\omega_e-\omega_g$. Associated raising and lowering operators are defined by $\sigma^+ =\ket{e}\bra{g}$ and $\sigma^-=\ket{g}\bra{e}$. The electromagnetic bath is described by creation and annihilation operators $a_{{\bf k}\lambda}^\dagger,~a_{{\bf k}\lambda}$ for a single photon with momentum ${\bf k}$ and polarisation $\lambda$. The photon frequency is denoted $\omega_k = |{\bf k}|$.

The energy of the dipole-field system is given by a Hamiltonian of the form $H=H_0+V$, where 
\begin{align}\label{h0}
H_0 = \omega_0\sigma^+\sigma^- + \sum_{{\bf k}\lambda} \omega_{k} a_{{\bf k}\lambda}^\dagger a_{{\bf k}\lambda},
\end{align}
defines the free (unperturbed) Hamiltonian and $V$ denotes the interaction Hamiltonian. Gauge-freedom within the microscopic description results in the freedom to choose a number of possible interaction Hamiltonians. We define the generalised-gauge transformation \cite{stokes_extending_2012}
\begin{align}\label{R}
R_{\{\alpha_k\}}:=\exp\left[{\hat {\bf d}}\cdot {\bf A}_{\{\alpha_k\}}({\bf 0}) \right],\qquad {\bf A}_{\{\alpha_k\}}({\bf x}) = \sum_{{\bf k}\lambda } \left({ \frac{1}{2\omega_k L^3}}\right)^{1 \over 2}  \alpha_k  {\bf e}_{{\bf k}\lambda }  a_{{\bf k}\lambda } e^{i{\bf k}\cdot {\bf x}} + {\rm H.c.}
\end{align}
In this expression the $\alpha_k$ are real and dimensionless, the ${\bf e}_{{\bf k}\lambda},~\lambda=1,2$ are mutually orthogonal polarisation unit vectors, which are both orthogonal to ${\bf k}$, $L^3$ is the volume of the assumed fictitious quantisation cavity, and ${\hat {\bf d}}$ denotes the dipole moment operator. By making the two-level approximation {\em after} having transformed the Coulomb gauge Hamiltonian using the unitary operator $R_{\{\alpha_k\}}$ we obtain the Hamiltonian $H=H_0+V$ where \cite{stokes_extending_2012}
\begin{align}\label{ham3}
V = \left[\sum_{{\bf k}\lambda } g_{{\bf k} \lambda} \, \sigma^+ \left( u_k^+ \, a_{{\bf k}\lambda }^\dagger + u_k^- \, a_{{\bf k}\lambda } \right) +{\rm H.c}\right]+V^{(2)}
\end{align}
and
\begin{align}\label{V2}
 V^{(2)}=\sum_{{\bf k}\lambda} {1\over 2L^3} \, \alpha_k^2 \, |{\bf e}_{{\bf k}\lambda }\cdot {\bf d} |^2+{e^2 \over 2m} \, {\tilde {\bf A}}({\bf 0})^2.
\end{align}
The term $V^{(2)}$ is a self-energy term, which does not act within the two-level dipole Hilbert space and which depends on the field
\begin{align}\label{At}
 {\tilde {\bf A}}({\bf x}) = \sum_{{\bf k}\lambda } \left({ \frac{1}{2\omega L^3}}\right)^{1 \over 2} (1-\alpha_k){\bf e}_{{\bf k}\lambda } \, a_{{\bf k}\lambda}e^{i{\bf k}\cdot {\bf x}} + {\rm H.c.}.
\end{align}
The coupling constant $g_{{\bf k} \lambda}$ and the (real) coefficients $u_k^\pm$ are defined as
\begin{align} \label{u}
g_{{\bf k} \lambda}=  -i \left({\omega_0 \over 2 L^3}\right)^{1\over 2} \, {\bf  e}_{{\bf k}\lambda }\cdot{\bf d},\qquad
u_k^\pm =  (1-\alpha_k) \left( {\omega_0 \over \omega_k} \right)^{1/2} \mp \alpha_k \left( {\omega_k \over \omega_0} \right)^{1/2}
\end{align}
where ${\bf d}$ and $\omega_0$ denote the two-level transition dipole moment and transition frequency respectively. The real numbers $\alpha_k$ can be chosen arbitrarily. Choosing $\alpha_k=0$ yields the Coulomb-gauge Hamiltonian while choosing $\alpha_k=1$ yields the multipolar-gauge Hamiltonian.  Letting $\alpha_k=1$ in Eq.~(\ref{R}) yields the well-known Power-Zienau-Woolley (PZW) transformation that relates the Coulomb and multipolar gauges. While the relation between the Coulomb and multipolar gauge has been discussed extensively \cite{milonni_natural_1989,woolley_gauge_1998,power_time_1999,power_time_1999-1,stokes_gauge_2013}, the PZW transformation is in fact a special case of a broader class of unitary gauge-fixing transformations \cite{stokes_gauge_2013,woolley_charged_1999,stokes_noncovariant_2012}. More generally still, the freedom to choose the $\alpha_k$ within the canonical transformation (\ref{R}) implies redundancy within our mathematical description and is henceforth referred to as generalised gauge-freedom. A third special case of Eq.~(\ref{ham3}) is afforded by making the choice $\alpha_k=\omega_0/(\omega_0+\omega_k)$, which specifies a symmetric mixture of Coulomb and multipolar couplings. This representation has proved useful in both photo-detection theory \cite{stokes_gauge_2013,drummond_unifying_1987} and open quantum systems theory \cite{stokes_extending_2012}, because within this representation $u_k^+\equiv 0$. The counter-rotating terms in the linear dipole-field interaction term $V-V^{(2)}$ are thereby eliminated without use of the rotating-wave approximation.

Given the above arbitrary generalised-gauge description, it is clear that arbitrary matrix elements $M_{fi}(t)=\bra{f}M(t)\ket{i}$ between eigenstates $\ket{f}$ and $\ket{i}$ of $H_0$ will not be the same when the evolution of the operator $M$ is determined by different total Hamiltonians $H$ and $H'$, that have been obtained by making different choices of $\alpha_k$ in Eq.~(\ref{ham3}). In contrast on-energy-shell QED $S$-matrix elements are necessarily the same for two interaction Hamiltonians $V$ and $V'$, constrained such that the corresponding total Hamiltonians are related by a unitary transformation $e^{iT}$ as \cite{craig_molecular_1984,cohen-tannoudji_photons_1989,woolley_gauge_1998}
\begin{align}\label{gic}
H=H_0+V,\qquad H'=e^{iT}He^{-iT}=H_0+V'.
\end{align}
For gauge-invariance to hold the unperturbed Hamiltonian $H_0$ must be identified as the same operator before and after the transformation by $e^{iT}$. Note however, that the unperturbed Hamiltonian $H_0$ given in Eq.~(\ref{h0}) does not commute with the unitary transformation $R_{\{\alpha_k\}}$ given in Eq.~(\ref{R}) meaning that this $H_0$ represents a different physical observable depending on the choice of interaction. Despite this, $S$-matrix elements based on the partition $H=H_0+V$ are invariant, because $H_0$ in Eq.~(\ref{h0}) does not explicitly depend on the $\alpha_k$ and is therefore the same for each different choice of generalised-gauge.

Having determined the conditions under which QED matrix elements are gauge-invariant we now turn our attention to deriving a master equation describing the two-level dipole within the radiation field. The conventional derivation of the second order quantum optical master equation, as found in Ref.~\cite{breuer_theory_2002} for example, does not at any point involve self-energy contributions due to the $V^{(2)}$ term within the interaction $V$. In general however, this term does contribute to dipole level-shifts, as is shown in appendix \ref{ap1}. In the general case that the temperature of the radiation field is arbitrary, the self-energy contributions from $V^{(2)}$ can be incorporated into the master equation by defining the Hamiltonian
\begin{align}\label{hdt}
{\tilde H}_d= (\omega_0 + \delta_e^{(2)}-\delta_g^{(2)})\sigma^+\sigma^-
\end{align}
where the excited and ground state self-energy shifts are defined as
\begin{align}\label{ses}
\delta_n^{(2)} = {\rm tr}(V\ket{n}\bra{n}\otimes \rho_F^{\rm eq})={\rm tr}(V^{(2)}\ket{n}\bra{n}\otimes \rho_F^{\rm eq}),
\end{align}
in which $n=e,g$ and $\rho_F^{\rm eq}(\beta)=e^{-\beta\sum_{{\bf k}\lambda} \omega_k a_{{\bf k}\lambda}^\dagger a_{{\bf k}\lambda}}/{\rm tr}(e^{-\beta\sum_{{\bf k}\lambda} \omega_k a_{{\bf k}\lambda}^\dagger a_{{\bf k}\lambda}})$ with $\beta$ the inverse temperature of the radiation field. Since $V^{(2)}$ is gauge-dependent we cannot include the self-energy shifts within the unperturbed Hamiltonian $H_0$ without ruining the gauge-invariance of any $S$-matrix elements obtained using the unperturbed states. Instead we replace the free system Hamiltonian in the usual Born-Markov master equation with the shifted Hamiltonian ${\tilde H}_d$ directly to obtain the second order master equation
\begin{align}\label{BM3}
{\dot \rho} =-i[{\tilde H}_d,\rho]-e^{-iH_0 t}\int^\infty_0 ds\, {\rm tr}_F\big[V_I(t),[V_I(t-s),\rho_I(t)\otimes\rho_F^{\rm eq}]\big]e^{iH_0 t}.
\end{align}
This master equation automatically includes the level shifts due to $V^{(2)}$ within the unitary evolution part, but the rest of the master equation is expressed in terms of the original partition $H=H_0+V$. Using this partition where $H_0$ and $V$ are given in Eqs.~(\ref{h0}) and (\ref{ham3}) respectively, Eq.~(\ref{BM3}) yields the $\alpha_k$-independent result
\begin{align}\label{me1}
{\dot \rho} =& -i[{\tilde \omega}_0\sigma^+\sigma^-,\rho] +\gamma(N+1)\left(\sigma^-\rho\sigma^+ -{1\over 2}\{\sigma^+\sigma^-,\rho\}\right)+ \gamma N\left(\sigma^+\rho\sigma^- - {1\over 2}\{\sigma^-\sigma^+,\rho\}\right).
\end{align}
Here $N=1/(e^{\beta\omega_0}-1)$, ${\tilde \omega}_0=\omega_0+\Delta$ and
\begin{align}\label{rates1}
\gamma &= \, 2\pi \sum_{{\bf k}\lambda} |\bra{{\bf k}\lambda,g}V\ket{0,e}|^2\delta(\omega_k-\omega_0) = {\omega_0^3|{\bf d}|^2 \over 3\pi}\nonumber \\ \Delta &= \int {d^3k \over (2\pi)^3}\sum_\lambda |{\bf e}_\lambda({\bf k})\cdot {\bf d}|^2(1+2N_k){\omega_0^3 \over \omega_k (\omega_0^2-\omega_k^2)}
\end{align}
where the continuum limit for wavevectors ${\bf k}$ has been applied and $N_k=1/(e^{\beta\omega_k}-1)$. Further details of the calculations leading to the final result for $\Delta$ in Eq.~(\ref{rates1}) are given in Appendix \ref{ap1}. We note that for $N_k=0$ we have
\begin{align}\label{delta0temp}
\Delta &= \,{\tilde \omega}_0-\omega_0  = \bra{0,e}V\ket{0,e} + \sum_{{\bf k}\lambda} {|\bra{0,e}V\ket{{\bf k}\lambda,g}|^2 \over \omega_0-\omega_k} -\bra{0,g}V\ket{0,g} +\sum_{{\bf k}\lambda} {|\bra{0,g}V\ket{{\bf k}\lambda,e}|^2 \over \omega_0+\omega_k}  \nonumber \\ &= \int {d^3k\over  (2\pi)^3} \sum_\lambda |{\bf e}_{{\bf k}\lambda}\cdot {\bf d}|^2 {\omega_0^3 \over  \omega_k(\omega_0^2-\omega_k^2)},
\end{align}
The $\alpha_k$-independence (generalised gauge-invariance) of the master equation (\ref{me1}) can be understood by noting that the spontaneous emission rate $\gamma$ and level-shift $\Delta$ in Eq.~(\ref{delta0temp}) are gauge-invariant QED matrix elements that can be obtained directly using second order perturbation theory.

In summary, we have shown that the master equation obtained from different, unitarily equivalent microscopic Hamiltonians is the same provided it depends only on $S$-matrix elements. $S$-matrix elements are invariant if the bare Hamiltonian $H_0$ is kept the same for each choice of total Hamiltonian. In what follows this will be seen to be significant for the derivation of the master equation describing two strongly coupled dipoles.

\subsection{Arbitrary gauge derivation of the standard two-dipole master equation}

Let us now turn our attention to obtaining the analogous result to Eq.~(\ref{me1}) for the case of two identical interacting dipoles at positions ${\bf R}_1$ and ${\bf R}_2$ within a common radiation reservoir. The transition dipole moments and transition frequencies of the dipoles are independent of the dipole label and are denoted ${\bf d}$ and $\omega_0$ respectively. The wavelength corresponding to $\omega_0$ is denoted with $\lambda_0$. An important quantity in the two-dipole system dynamics is the inter-dipole separation $R=|{\bf R}_2-{\bf R}_1|$. In terms of $R$ we can identify in the usual way three distinct parameter regimes: $R \ll \lambda_0$ is the near zone in which $R^{-3}$-dependent terms dominate, $R\sim \lambda_0$ is the intermediate zone in which $R^{-2}$-dependent terms may become significant, and $R \gg \lambda_0$ is the far-zone (radiation zone) in which $R^{-1}$-dependent terms dominate. We give a general derivation of the standard two dipole master equation, in which the gauge freedom within the microscopic Hamiltonian is left open throughout. This reveals limitations within the conventional derivation using the multipolar gauge. To begin we define the two-dipole generalised Power-Zienau-Woolley gauge transformation by \cite{drummond_unifying_1987,stokes_extending_2012}
\begin{align}
R_{\{\alpha_k\}}:=\exp\left(i\sum_{\mu=1}^2 {\bf d}_\mu \cdot {\bf A}_{\{\alpha_k\}}({\bf R}_\mu) \right)
\end{align}
where ${\bf d}_\mu$ denotes the dipole moment operator of the $\mu$'th dipole. We now transform the dipole approximated Coulomb gauge Hamiltonian using $R_{\{\alpha_k\}}$ and afterwards make the two-level approximation for each dipole. This implies that ${\bf d}_\mu = {\bf d}(\sigma^+_\mu+\sigma^-_\mu)$, and that the Hamiltonian can be written $H=H_0+V$ with
\begin{align}\label{h02}
H_0=\sum_{\mu=1}^2 \omega_0\sigma_\mu^+\sigma_\mu^-+ \sum_{{\bf k}\lambda } \omega_k a_{{\bf k}\lambda }^\dagger a_{{\bf k}\lambda }
\end{align}
and
\begin{align}\label{intgen}
V= \left[\sum_{{\bf k}\lambda } \sum_{\mu=1}^2 g_{{\bf k} \lambda} \sigma_\mu^+ \left( u_k^+ \, a_{{\bf k}\lambda }^\dagger e^{-i{\bf k}\cdot {\bf R}_\mu} + u_k^- \, a_{{\bf k}\lambda}e^{i{\bf k}\cdot{\bf R}_\mu} \right) +{\rm H.c} \right]+ V^{(2)}_{\{\alpha_k\}} + C_{\{\alpha_k\}}.
\end{align}
Analogously to the single-dipole case the first line in Eq.~(\ref{intgen}) defines a linear dipole-field interaction component while the term $V^{(2)}_{\{\alpha_k\}}$ consists of self-energy contributions for each dipole and the radiation field;
\begin{align}
V^{(2)}_{\{\alpha_k\}} = \sum_{{\bf k}\lambda} {1\over L^3} \, \alpha_k^2 \, |{\bf e}_{{\bf k}\lambda }\cdot {\bf d} |^2+\sum_{\mu=1}^2 {e^2 \over 2m} \, {\tilde {\bf A}}({\bf R}_\mu)^2,
\end{align}
where $m$ is the dipole mass. Due to the two-level approximation the first term is proportional to the identity, while the second term is a radiation self-energy term. The field ${\tilde A}$ is defined as in the singe-dipole case by Eq.~(\ref{At}). The final term in Eq.~(\ref{intgen}) $C_{\{\alpha_k\}}$, has no analog in the single-dipole Hamiltonian. This term gives a static Coulomb-like interaction between the dipoles, which is independent of the field;
\begin{align}\label{Calpha}
C_{\{\alpha_k\}} = \sum_{{\bf k}\lambda} {1\over L^3} \, (\alpha_k^2-1) \, |{\bf e}_{{\bf k}\lambda }\cdot {\bf d} |^2e^{i{\bf k}\cdot {\bf R}}\sigma^x_1\sigma^x_2,
\end{align}
where $\sigma_\mu^x = \sigma_\mu^++\sigma_\mu^-$. In the Coulomb gauge ($\alpha_k=0$) the term $C_{\{\alpha_k\}}$ reduces to the usual dipole-dipole Coulomb interaction. In the multipolar gauge $C_{\{\alpha_k\}}$ vanishes, and the interaction in Eq.~(\ref{intgen}) therefore reduces to a sum of interaction terms for each dipole. 

It is important to note that as in the single-dipole case $R_{\{\alpha_k\}}$ does not commute with $H_0$ given in Eq.~(\ref{h02}) implying that $H_0$ represents a different physical observable for each choice of $\alpha_k$. More generally, since $R_{\{\alpha_k\}}$ is a non-local transformation, which mixes material and transverse field degrees of freedom, the canonical material and field operators are different for each choice of $\alpha_k$. This implies that the master equation for the dipoles will generally be different for each choice of $\alpha_k$. We can, however, obtain a gauge-invariant result by ensuring that the master equation depends only on gauge-invariant $S$-matrix elements. These matrix elements are gauge-invariant {\em despite} the implicit difference in the material and field degrees of freedom within each generalised gauge. Usually the two-dipole master equation is derived using the specific choice $\alpha_k=1$ (multipolar gauge) for which the direct Coulomb-like coupling $C_{\{\alpha_k\}}$ vanishes identically. To obtain the same master equation for any other choice of $\alpha_k\neq 1$, we must include $C_{\{\alpha_k\}}$ within the interaction Hamiltonian $V$. The reason is that $H_0$ must be identified as the same operator for each choice of $\alpha_k$ in order that the gauge-invariance of the associated $S$-matrix holds.

We now proceed with a direct demonstration that the standard two-dipole master equation can indeed be obtained for any other choice of $\alpha_k$, provided $C_{\{\alpha_k\}}$ is kept within the interaction Hamiltonian $V$. To do this we substitute the interaction Hamiltonian in Eq.~(\ref{intgen}) into the second order Born-Markov master equation in the interaction picture with respect to $H_0$, which is given by
\begin{align}\label{BM}
{\dot \rho}_I(t) =-i{\rm tr}_F[V_I(t),\rho_I(t)\otimes\rho_F^{\rm eq}]
-\int^\infty_0 ds\, {\rm tr}_F\big[V_I(t),[V_I(t-s),\rho_I(t)\otimes\rho_F^{\rm eq}]\big],
\end{align}
where $V_I(t)$ denotes the interaction Hamiltonian in the interaction picture and $\rho_I(t)$ denotes the interaction picture state of the two dipoles. We retain contributions up to order $e^2$ and perform a further secular approximation, which neglects terms oscillating with twice the transition frequency $\omega_0$.  Transforming back to the Schr\"odinger picture and including the single-dipole self-energy contributions as in Eq.~(\ref{BM3}), we arrive after lengthy but straightforward manipulations at the final $\alpha_k$-independent result
\begin{align}\label{me2}
{\dot \rho} =  -i{\tilde \omega}_0\sum_{\mu=1}^2 [\sigma^+_\mu\sigma^-_\mu,\rho]  -i\Delta_{12}\sum_{\mu\neq\nu}^2 [\sigma_\mu^+\sigma_\nu^-, \rho] + \sum_{\mu,\nu=1}^2 \gamma_{\mu\nu}\Bigg[ &(N+1)\left(\sigma_\mu^-\rho \sigma_\nu^+-{1\over 2}\{\sigma_\mu^+\sigma_\nu^-,\rho\} \right)\nonumber \\ &+ N\left(\sigma_\mu^+\rho \sigma_\nu^--{1\over 2}\{\sigma_\mu^-\sigma_\nu^+,\rho\}\right)\Bigg].
\end{align}
This equation is identical in form to the standard two-dipole master equation, which can be found in Ref.~\cite{agarwal_quantum_2012} for example. The coefficients within the master equation are as follows. The decay rates $\gamma_{\mu\nu}$ are given by
\begin{align}\label{coeffs1} 
\gamma_{\mu\mu}&=\gamma,\nonumber \\ 
\gamma_{12}&=\, \gamma_{21}=2\pi \sum_{{\bf k}\lambda} \bra{0,e,g}V\ket{{\bf k}\lambda,g,g}\bra{{\bf k}\lambda,g,g}V\ket{0,g,e}\delta(\omega_0 - \omega_k)= d_i d_j \tau_{ij}(\omega_0, R)
\end{align}
where
\begin{align}\label{tau}
\tau_{ij}(\omega_0, R) = {\omega^3_0\over 2\pi}\Bigg((\delta_{ij}-{\hat R}_i {\hat R}_j){\sin \omega_0 R \over \omega_0 R} +(\delta_{ij}-3{\hat R}_i{\hat R}_j)\left[{\cos\omega_0R\over(\omega_0 R)^2}-{\sin\omega_0 R\over (\omega_0R)^3} \right]\Bigg).
\end{align}
In Eq.~(\ref{coeffs1}) and throughout we denote spatial components with Latin indices and adopt the convention that repeated Latin indices are summed. The quantity $\gamma_{12}$ denotes an $R$-dependent collective decay rate. The third equality in Eq.~(\ref{coeffs1}) wherein $\gamma_{12}$ has been expressed as a matrix element involving $V$ makes the reason for the $\alpha_k$-independence of this rate clear. We now turn our attention to the master equation shifts $\Delta={\tilde \omega}_0-\omega_0$ and $\Delta_{12}$. The single-dipole shift $\Delta$ includes all self-energy contributions, which have been dealt with in the same way as for the single-dipole master equation [cf Eq.~(\ref{BM3})]. The shift $\Delta$ is therefore as in Eq.~(\ref{rates1}). Details of the calculation of the joint shift $\Delta_{12}$ are given in appendix \ref{ap2} with the final result being
\begin{align}\label{jsfin}
\Delta_{12}= \sum_n {\bra{f}V\ket{n}\bra{n}V\ket{i}\over \omega_i - \omega_n}= \int {d^3k\over (2\pi)^3} \sum_\lambda |{\bf e}_\lambda({\bf k})\cdot {\bf d}|^2e^{i{\bf k}\cdot {\bf R}} {\omega_k^2 \over \omega_0^2-\omega_k^2} = d_i d_j V_{ij}(\omega_0, R),
\end{align}
where $\ket{f}= \ket{g,e,0}$, $\ket{i}=\ket{e,g,0}$ and $\ket{n}$ are eigenstates of $H_0$ and
\begin{align}\label{Vij}
V_{ij}(\omega_0, R)=-{\omega_0^3\over 4\pi}\Bigg((\delta_{ij}-{\hat R}_i {\hat R}_j){\cos \omega_0 R \over \omega_0 R} -(\delta_{ij}-3{\hat R}_i{\hat R}_j)\left[{\sin\omega_0R\over(\omega_0 R)^2}+{\cos\omega_0 R\over (\omega_0R)^3} \right]\Bigg).
\end{align}
As indicated by the second equality in Eq.~(\ref{jsfin}) $\Delta_{12}$ is nothing but the well-known gauge-invariant QED matrix-element describing resonant energy-transfer. 

We have therefore obtained the standard result, Eq.~(\ref{me2}), without ever making a concrete choice for the $\alpha_k$. In order that the standard result is obtained the direct Coulomb-like interaction $C_{\{\alpha_k\}}$ must be kept within the interaction Hamiltonian $V$. This ensures that for all $\alpha_k$ the unperturbed Hamiltonian $H_0$ is that used in conventional derivations wherein $\alpha_k=1$. The $S$-matrix elements involving $C_{\{\alpha_k\}}$ that appear as coefficients in the master equation are then $\alpha_k$-independent and are the same as those obtained in the conventional derivation. Our derivation makes it clear that when $C_{\{\alpha_k\}}$ is sufficiently strong compared with the linear dipole-field coupling term, its inclusion within the interaction Hamiltonian rather than $H_0$ may be ill-justified. The standard master equation may therefore be inaccurate in such regimes. This fact is obscured within conventional derivations that use the multipolar gauge $\alpha_k=1$, because in this gauge $C_{\{\alpha_k\}}$ vanishes identically. However, one typically still assumes weak-coupling to the radiation field in the multipolar gauge, and this leads to the standard master equation (\ref{me2}).  If instead we adopt the Coulomb gauge $\alpha_k=0$ we obtain the static dipole-dipole interaction $C_{\{0\}}=C\sigma_1^x\sigma_2^x$, where in the mode continuum limit
\begin{align}\label{Cdip}
C = {d_i d_j\over 4\pi R^3}(\delta_{ij}-3{\hat R}_i{\hat R}_j).
\end{align}
This quantity coincides with the near-field limit of the resonant energy transfer element $\Delta_{12}$ given in Eq.~(\ref{jsfin}). In the near-field regime $R/\lambda \ll 1$, $C_{\{0\}}$ may be too strong to be kept within the purportedly weak perturbation $V$ and the standard master equation, which only results when one treats $C_{\{0\}}$ as a weak perturbation, should then break down. This will be discussed in more detail in the following section.

\section{Corrections to the standard master equation}\label{2}

\subsection{Derivation of an alternative master equation}

In the near-field regime $R/\lambda_0\ll 1$ the rate of spontaneous emission into the transverse field is much smaller than the direct dipolar coupling; $C/\gamma \gg 1$. Moreover, for a system of closely spaced Rydberg atoms, the electrostatic Coulomb interaction may be such that $C\sim \omega_0$. For example, given a Rydberg state with principal quantum number $n=50$ we can estimate the maximum associated dipole moment as $(3/2)n^2 a_0 e \sim 10^{-26}{\rm Cm}$ where $a_0$ is the Bohr radius and $e$ the electronic charge. For a $1{\rm \mu m}$ separation, which is approximately equal to $10n^2a_0$, the electrostatic dipole interaction $C/\omega_0\sim1$ for $\omega_0$ corresponding to a microwave frequency. In such situations it is not clear that the Coulomb interaction can be included within the perturbation $V$ with the coupling to the transverse field. In the multipolar gauge where no direct Coulomb interaction is explicit the same physical interaction is mediated by the low frequency transverse modes, which must be handled carefully. A procedure which separates out these modes should ultimately result in a separation of the Coulomb interaction, which is of course already explicit within the Coulomb gauge. We remark that when considering realistic Rydberg atomic systems within the strong dipole-dipole coupling regime the validity of the two-level model should also be considered. However, moving beyond the two-level approximation is beyond the scope of this paper. Our aim is to consider general atomic, molecular and condensed matter systems strongly-coupled by dipole-dipole interactions for which two-level models are typically used \cite{jaksch_fast_2000,westermann_dynamics_2006,agarwal_quantum_2012}. %In reference \cite{jaksch_fast_2000} for example a system of two two-level dipoles strongly coupled by dipole-dipole interactions is considered as a potential implementation for quantum gates.
Retaining the two-level model for each dipole allows us to succinctly compare with existing literature and thereby determine the relative difference produced by our non-perturbative treatment of dipole-dipole interactions.

In the Coulomb gauge the interaction Hamiltonian $V$ coupling to the transverse radiation field is
\begin{align}\label{cgint}
V=&\sum_{\mu=1}^2 \omega_0\sigma_\mu^y{\bf d}\cdot {\bf A}({\bf R}_\mu)+\sum_{\mu=1}^2 {e^2\over 2m}{\bf A}({\bf R}_\mu)^2
\end{align}
with $\sigma^y_\mu = -i(\sigma^+_\mu-\sigma^-_\mu)$ and
\begin{align}\label{A}
{\bf A}({\bf x}) = \sum_{{\bf k}\lambda} \sqrt{{1 \over 2\omega_k L^3}} {\bf e}_{{\bf k}\lambda} a_{{\bf k}\lambda}^\dagger e^{-i{\bf k}\cdot{\bf x}} +{\rm H.c.}\, .
\end{align}
The contribution of the transverse field to $\Delta_{12}$ in Eq. (\ref{coeffs1}) is found using Eq. (\ref{cgint}) to be
\begin{align}\label{d12lim}
{\tilde \Delta}_{12} =& \sum_n {\bra{0,e,g}V\ket{n}\bra{n}V\ket{0,g,e}\over \omega_0 - \omega_n}\nonumber \\
 =&  -{\omega_0^3 d_i d_j\over 4\pi}\Bigg((\delta_{ij}-{\hat R}_i {\hat R}_j){\cos \omega_0 R \over \omega_0 R} -(\delta_{ij}-3{\hat R}_i{\hat R}_j)\left[{\sin \omega_0R\over(\omega_0 R)^2}-{1-\cos \omega_0 R \over (\omega_0R)^3} \right]\Bigg)\nonumber \\ =& \, \Delta_{12}-C.
\end{align}

When the contribution $C=\bra{0,e,g}C_{\{0\}}\ket{0,g,e}$ resulting from the direct Coulomb interaction is added to ${\tilde \Delta}_{12}$ the fully retarded result $\Delta_{12}$ is obtained. The two matrix elements $\Delta_{12}$ and ${\tilde \Delta}_{12}$ therefore only differ in their near-field components, which vary as $R^{-3}$ and which we denote by $\Delta_{12}^{\rm nf}$ and ${\tilde \Delta}_{12}^{\rm nf}$, respectively. According to Eqs.~(\ref{jsfin}) and (\ref{d12lim}) the components $\Delta_{12}^{\rm nf}$ and ${\tilde \Delta}_{12}^{\rm nf}$ dominate at low frequencies $\omega_0$. Since $\Delta_{12}$ is evaluated at resonance $\omega_k=\omega_0$, it follows that within the multipolar-gauge the low $\omega_k$ modes within the system-reservoir coupling give rise to a strong dipole-dipole interaction in the form of $\Delta_{12}^{\rm nf}\approx C$. In such regimes the multipolar interaction Hamiltonian cannot be classed as a weak perturbation. On the other hand, the matrix element ${\tilde \Delta}_{12}$ is obtained using the Coulomb gauge interaction in Eq.~(\ref{cgint}), and is such that ${\tilde \Delta}_{12}^{\rm nf}=\Delta_{12}^{\rm nf}-C\approx 0$. Within the Coulomb gauge the interaction equivalent to the low frequency part of the multipolar gauge system-reservoir coupling is a direct dipole-dipole Coulomb interaction $C_{\{0\}}$. This appears explicitly in the Hamiltonian, but  has not been included within Eq.~(\ref{cgint}), which therefore represents a genuinely weak perturbation.

The collective decay rate $\gamma_{12}$ as given in Eq.~(\ref{coeffs1}) does not involve the direct Coulomb interaction $C_{\{0\}}$ in any way, and can be obtained from the transverse field interaction in Eq.~(\ref{cgint}) or from the multipolar interaction. Crucially, in the near-field regime  $R/\lambda_0\ll 1$ the terms $\gamma,~\gamma_{12}$ and ${\tilde \Delta}_{12}$, which result from the interaction in Eq.~(\ref{cgint}), are several orders of magnitude smaller than the direct electrostatic coupling $C$. Motivated by the discussion above, we include the Coulomb interaction within the unperturbed Hamiltonian, but continue to treat the interaction with the transverse field as a weak perturbation. This gives rise to a master equation depending on different $S$-matrix elements.

The unperturbed Hamiltonian $H_0=H_d+H_F$ is defined by
\begin{align}\label{free2}
H_d= \sum_{\mu=1}^2 \omega_0\sigma_\mu^+\sigma_\mu^- + C\sigma_1^x\sigma_2^x, \qquad
H_F=\sum_{{\bf k}\lambda}\omega_k a_{{\bf k}\lambda}^\dagger a_{{\bf k}\lambda},
\end{align}
where $C\in {\mathbb R}$ is given by Eq.~(\ref{Cdip}). The corresponding interaction Hamiltonian is then given in Eq.~(\ref{cgint}).
%while the interaction is given in Eq.~(\ref{cgint}). In Eq. (\ref{free2}) $C_{\{0\}}=C\sigma^x_1\sigma^x_2$ is defined as in Eq. (\ref{Calpha}), and in the mode continuum limit $C\in {\mathbb R}$ is given by Eq. (\ref{Cdip}). 
We begin by diagonalising $H_d$ as
\begin{align}\label{hd}
H_d = \sum_{n=1}^4 \epsilon_n\ket{\epsilon_n}\bra{\epsilon_n}
\end{align}
where
\begin{align}
\epsilon_1 = \omega_0-\eta,\qquad \epsilon_2=\omega_0-C \qquad \epsilon_3= \omega_0+C,\qquad \epsilon_4 = \omega_0+\eta,
\end{align}
and
\begin{align}\label{estates}
\ket{\epsilon_1} &= {1\over \sqrt{C^2+(\omega_0+\eta)^2}}\left([\omega_0+\eta]\ket{g,g}-C\ket{e,e}\right),\qquad
\ket{\epsilon_2} = {1\over \sqrt{2}}\left(\ket{e,g}-\ket{g,e}\right),\nonumber \\
\ket{\epsilon_3} &= {1\over \sqrt{2}}\left(\ket{e,g}+\ket{g,e}\right),\qquad
\ket{\epsilon_4} = {1\over \sqrt{C^2+(\omega_0-\eta)^2}}\left(C\ket{e,e}-[\omega_0-\eta]\ket{g,g}\right),
\end{align}
with $\eta=\sqrt{\omega_0^2+C^2}$. Next we move into the interaction picture with respect to $H_0$ and substitute the interaction picture interaction Hamiltonian into Eq.~(\ref{BM}). Moving back into the Schr\"odinger picture we eventually obtain
\begin{align}\label{me3}
{\dot \rho}=-i[H_d,\rho]+\sum_{\zeta,\zeta'=\pm\omega_{1,2}}\sum_{\mu,\nu=1}^2 \bigg[\Gamma_{\mu\nu}(\zeta)\big(A_{\nu\zeta}\rho A_{\mu\zeta'}^\dagger -A_{\mu\zeta'}^\dagger A_{\nu\zeta}\rho\big) + {\rm H.c.}\bigg].
\end{align}
Here, $\omega_1 = \eta-C$ and $\omega_2 = \eta+C$, while $A_{\mu(-\zeta)}=A_{\mu\zeta}^\dagger$ and $A_{\mu\zeta_n}\equiv A_{\mu n}~(n=1,2)$ with
\begin{align}
A_{11} &= a\ket{\epsilon_1}\bra{\epsilon_2}+b\ket{\epsilon_3}\bra{\epsilon_4},\qquad~~\,
A_{12} = c\ket{\epsilon_1}\bra{\epsilon_3}-d\ket{\epsilon_2}\bra{\epsilon_4},\nonumber \\
A_{21} &= -a\ket{\epsilon_1}\bra{\epsilon_2}+b\ket{\epsilon_3}\bra{\epsilon_4},\qquad
A_{22} = c\ket{\epsilon_1}\bra{\epsilon_3}+d\ket{\epsilon_2}\bra{\epsilon_4},
\end{align}
where
\begin{align}
a&={\omega_0+\eta-C \over \sqrt{2(C^2+[\omega_0+\eta]^2)}},\qquad b={\omega_0-\eta-C \over \sqrt{2(C^2+[\omega_0-\eta]^2)}},\nonumber\\
c&={\omega_0+\eta+C \over \sqrt{2(C^2+[\omega_0+\eta]^2)}},\qquad d={-\omega_0+\eta+C \over \sqrt{2(C^2+[\omega_0-\eta]^2)}}.\nonumber\\
\end{align}
The coefficients $\Gamma_{\mu\nu}(\omega)$ are defined by
\begin{align}
\Gamma_{\mu\nu}(\omega) = \omega_0^2 d_i d_j \int_0^\infty ds \,e^{i\omega s} \langle A_{I,i}({\bf R}_\mu,s)A_{I,j}({\bf R}_\nu,0)\rangle_{\beta},
\end{align}
where ${\bf A}_I({\bf x},t)$ denotes the field ${\bf A}({\bf x})$ in Eq.~(\ref{A}) once transformed into the interaction picture, and $\langle\cdot\rangle_\beta$ denotes the average with respect to the radiation thermal state at temperature $\beta^{-1}$. The $\Gamma_{\mu\nu}(\omega)$ are symmetric $\Gamma_{\mu\nu}(\omega)=\Gamma_{\nu\mu}(\omega)$ and can be written 
\begin{align}
\Gamma_{\mu\nu}(\omega)={1\over 2}\gamma_{\mu\nu}(\omega) +iS_{\mu\nu}(\omega)
\end{align}
where
\begin{align}\label{coeffs3}
\gamma_{\mu\mu}(\omega) & = (1+N)\gamma{\omega\over \omega_0}, \qquad
\gamma_{12}(\omega)= (1+N)d_id_j\tau_{ij}(\omega, R){\omega_0^2\over \omega^2},\nonumber \\
S_{\mu\mu}(\omega) &= {\gamma\over 2\pi}\int_0^\infty d\omega_k {\omega_k\over \omega_0}\left[{1+N_k\over \omega-\omega_k}+{N_k\over \omega+\omega_k}\right],\nonumber \\ 
S_{12}(\omega) &={d_id_j\over 2\pi}\hspace*{-1mm}\int_0^\infty d\omega_k \, \tau_{ij}(\omega_k,R){\omega_0^2 \over \omega_k^2} \hspace*{-0.5mm} \left[{1+N_k \over \omega-\omega_k}+{N_k\over \omega+\omega_k}\right],
\end{align}
with $N_k = 1/(e^{\beta\omega_k}-1)$. The frequency integrals in Eq.~(\ref{coeffs3}) are to be understood as principal values. 
The decay rates $\gamma_{\mu\nu}(\omega)$ in Eq.~(\ref{coeffs3}) coincide with those found in the standard master equation (\ref{me2}) when evaluated at $\omega_0$, though are here evaluated at the frequencies $\omega_{1,2}$. The quantities $S_{\mu\nu}$ are related to the shifts $\Delta$ and $\Delta_{12}$, %associated with the standard master equation (\ref{me2}) 
defined in Eqs.~(\ref{rates1}) and~(\ref{jsfin}) respectively, by
\begin{align}\label{dS}
&\Delta = S_{\mu\mu}(\omega_0)-S_{\mu\mu}(-\omega_0),
\qquad \Delta_{12}-C = {\tilde \Delta}_{12}=S_{12}(\omega_0)+S_{12}(-\omega_0).
\end{align}

In deriving Eq.~(\ref{me3}) we have not yet performed a secular approximation, in contrast to the derivation of Eq.~(\ref{me2}). However, naively applying a secular approximation that neglects off-diagonal terms for which $\zeta \neq \zeta'$ in the summand in Eq.~(\ref{me3}) would not be appropriate, because this would eliminate terms that are resonant in the limit $C\to 0$. Instead we perform a partial secular approximation which eliminates off-diagonal terms for which $\zeta$ and $\zeta'$ have opposite sign. These terms remain far off-resonance for all values of $C$. The resulting master equation is given by
\begin{align}\label{me4}
{\dot \rho}=&-i[H_d,\rho]\nonumber \\ &+\sum_{\zeta,\zeta'=\omega_{1,2}} \sum_{\mu,\nu=1}^2 \bigg[\Gamma_{\mu\nu}(\zeta)\big(A_{\nu\zeta}\rho A_{\mu\zeta'}^\dagger  -A_{\mu\zeta'}^\dagger A_{\nu\zeta}\rho\big) + \Gamma_{\mu\nu}(-\zeta)\big(A_{\nu\zeta}^\dagger \rho A_{\mu\zeta'} -A_{\mu\zeta'}A_{\nu\zeta}^\dagger\rho\big) + {\rm H.c.}\bigg].
\end{align}

We are now in a position to compare our master equation (\ref{me4}) with the usual result in~(\ref{me2}). In the limit $C\to 0$ we have $\eta\to \omega_0$ so that $\omega_{1,2} \to \omega_0$. The rates and shifts in Eq.~(\ref{coeffs3}) are then evaluated at $\omega_0$ within Eq.~(\ref{me4}). %Finally, in the limit $C\to 0$
Also, the Hamiltonian $H_d$ tends to the bare Hamiltonian $\omega_0(\sigma_1^+\sigma_1^-+\sigma_2^+\sigma_2^-)$, and furthermore we have that
\begin{align}\label{Alim}
\sum_{\zeta=\omega_{1,2}} A_{\mu\zeta} \to \sigma_\mu^-.
\end{align}
Thus, taking the limit $C\to 0$ in Eq.~(\ref{me4}) 
%and using eqs. (\ref{dS}) and (\ref{Alim}) 
one recovers Eq.~(\ref{me2}) with $\Delta_{12}$ replaced by ${\tilde \Delta}_{12}$ given in Eq.~(\ref{d12lim}). However, since ${\tilde \Delta}_{12} \to \Delta_{12}$ when $C\to 0$, Eqs.~(\ref{me4}) and (\ref{me2}) coincide in this limit. For finite $C$, Eq.~(\ref{me4}) offers separation-dependent corrections to the usual master equation and is the main result of this section.

\subsection{Discussion: gauge-invariance of the new master equation}

It is important to note that while our master equation (\ref{me4}) is generally different to the usual gauge-invariant result [Eq.~(\ref{me2})] there is no cause for concern regarding the issue of gauge-invariance. As we have shown the standard master equation can be obtained when $\alpha_k=0$ provided one uses a partitioning of the Hamiltonian in the form $H=H_0+V^{\rm usual}$ where $H_0$ is given by Eq.~(\ref{h02}) and
\begin{align}
V^{\rm usual}=\sum_{\mu=1}^2 \omega_0\sigma_\mu^y{\bf d}\cdot {\bf A}({\bf R}_\mu)+\sum_{\mu=1}^2 {e^2\over 2m}{\bf A}({\bf R}_\mu)^2
+C_{\{0\}}.
\end{align}
Our master equation (\ref{me4}) has also been obtained by choosing $\alpha_k=0$, but our derivation makes use of the different partitioning $H={\tilde H}_0+V$ where ${\tilde H}_0$ is defined as in Eq.~(\ref{free2}) and $V=V^{\rm usual}-C_{\{0\}}$ is defined as in Eq.~(\ref{cgint}). The two different partitionings of the same Hamiltonian yield two different second order master equations.

As we have shown the standard Born-Markov-secular master equation (\ref{me2}) can be obtained for any other choice of $\alpha_k$ provided that the unperturbed Hamiltonian is always defined as in Eq.~(\ref{h02}). Similarly a full secular approximation of our master equation (\ref{me4}) can also be obtained for any other choice of $\alpha_k$ provided the unperturbed Hamiltonian is always defined as in Eq. (\ref{free2}). We note further that the secular approximation is well justified within the near-field regime of interest $R\ll\lambda_0$. Let us consider for example the multipolar gauge obtained by choosing $\alpha_k= 1$. In order to achieve the appropriate partitioning of the multipolar Hamiltonian for derivation of our master equation one must add $C_{\{0\}}$ to the usual multipolar unperturbed Hamiltonian $H_0$ given in Eq.~(\ref{h02}), and simultaneously subtract $C_{\{0\}}$ from the usual multipolar interaction Hamiltonian. Using this repartitioning of the multipolar Hamiltonian the Born-Markov-secular master equation is found to coincide with our master equation (\ref{me4}) once a full secular approximation is performed within the latter. This derivation is however, more cumbersome than the Coulomb gauge derivation. Since the Coulomb energy is naturally explicit within the Coulomb gauge, the latter is the most natural gauge to choose for the purpose of including the relatively strong static interaction within the unperturbed Hamiltonian.

Any difference between the master equation (\ref{me4}) and the corresponding partially-secular result found using $\alpha_k\neq 0$ is contained entirely within non-secular contributions. These contributions are negligible within the regime of interest $R\ll \lambda_0$ and have only been retained within Eq. (\ref{me4}) to facilitate comparison with the standard result Eq.~(\ref{me2}). Moreover, in the far-field regime $R \gg \lambda_0$ the master equations (\ref{me2}) and (\ref{me4}) coincide, so the master equation (\ref{me4}) is also gauge-invariant within this regime.

\section{Solutions and emission spectrum}\label{3}

\subsection{Solutions}

\begin{figure}
\begin{minipage}{\columnwidth}
\begin{center}
\includegraphics{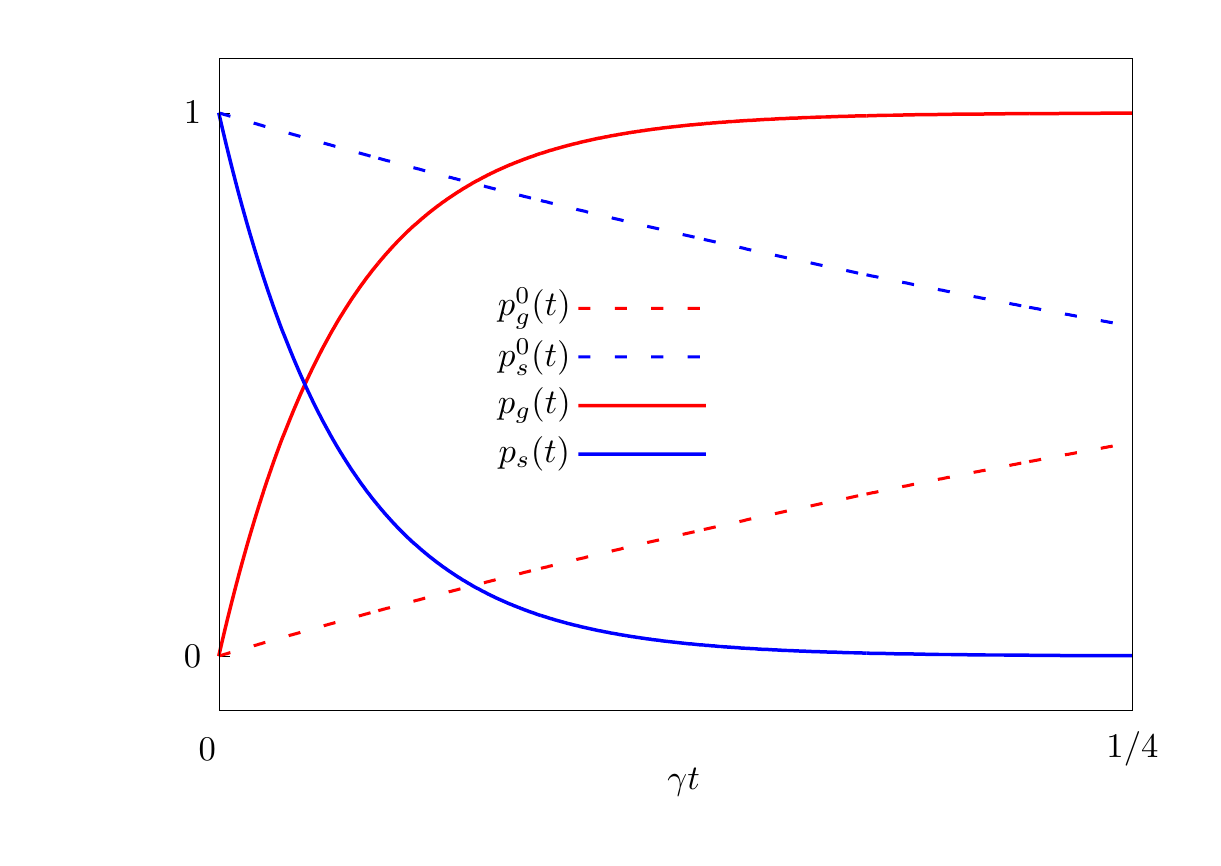}
\vspace*{-0.5cm}
\caption{The populations of the stationary state ($\ket{\epsilon_1}$ or $\ket{g,g}$) and symmetric state $\ket{\epsilon_3}$ are plotted as functions of $t$ for fixed separation $R=10 r_a$, where $r_a=n^2 a_0$ is a characteristic Rydberg atomic radius, with $n=50$ and $a_0$ the Bohr radius. We have assumed no thermal occupation of the field, $N=0$ and that the transition dipole moment ${\bf d}$ is orthogonal to the separation vector ${\bf R}$. The transition frequency is chosen in the microwave regime $\omega_0 =10^{10}$. We use $p_g$ and $p_s$ to denote the stationary and symmetric state populations respectively, and we use $p$ and $p^0$ to denote populations obtained from master equations (\ref{me4}) and (\ref{me2}), respectively. In the case of Eq. (\ref{me4}) the stationary state is $\ket{\epsilon_1}$ whereas in the case of Eq.~(\ref{me2}) the stationary state is simply $\ket{g,g}$. The initial condition chosen is $p_s=1$. The ground and symmetric state populations obtained from the master equation (\ref{me4}) crossover significantly earlier than those obtained from Eq.~(\ref{me2}).}\label{ft}
\end{center}
\end{minipage}
\end{figure}

\begin{figure}
\begin{minipage}{\columnwidth}
\begin{center}
\vspace*{-2cm}
\includegraphics{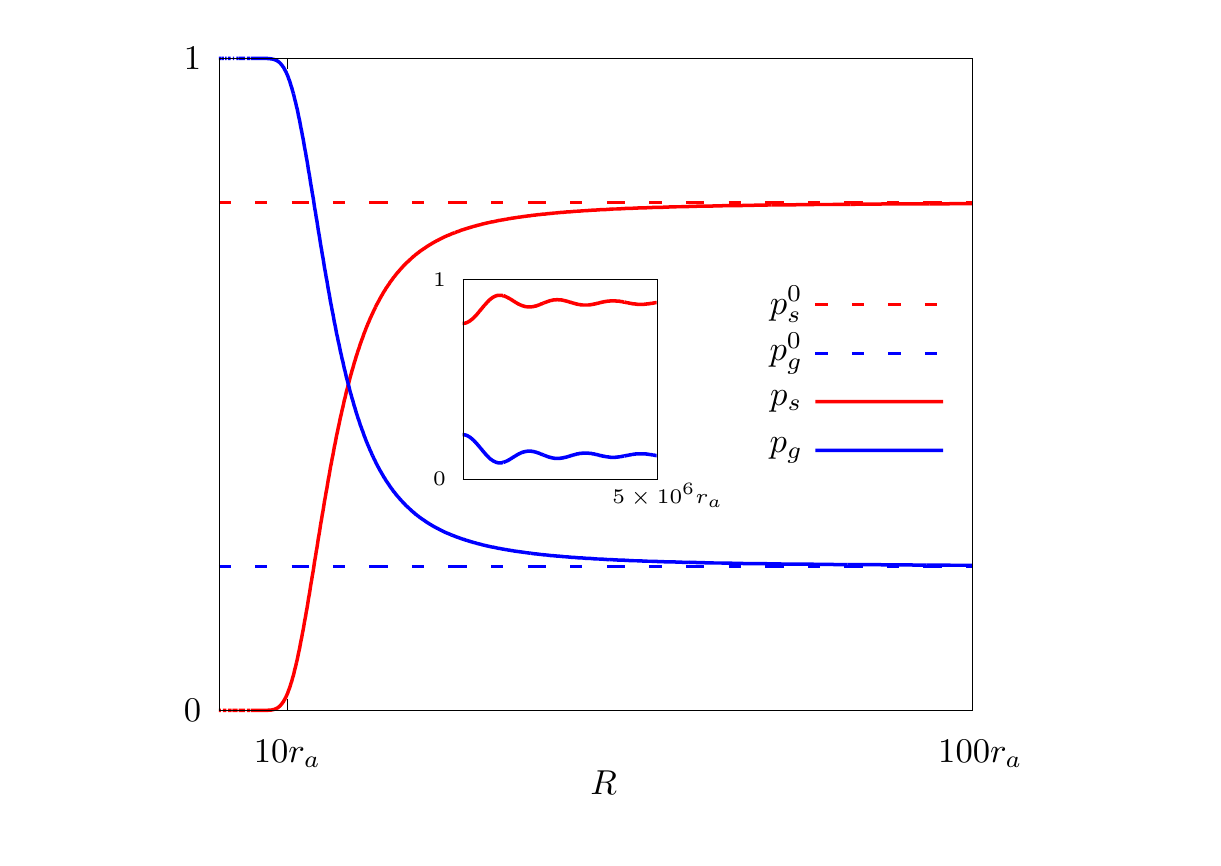}
\vspace*{-0.5cm}
\caption{The populations of the stationary state ($\ket{\epsilon_1}$ or $\ket{g,g}$) and symmetric state $\ket{\epsilon_3}$ are plotted as functions of $R$ for fixed time $t=1/8\gamma$. The remaining parameters are as in Fig.~\ref{ft}. For the separations considered the solutions to Eq.~(\ref{me2}) do not vary significantly, while the solutions to Eq.~(\ref{me4}) converge to those of Eq.~(\ref{me2}) only for larger values of $R$. The subplot shows these solutions over a much larger scale of separations up to $O(10^6 r_a)$, for which the solutions to Eqs.~(\ref{me2}) and (\ref{me4}) are indistinguishable.}\label{f1}
\end{center}
\end{minipage}
\end{figure}

For large inter-dipole separations $R\gg \lambda_0$ the master equations (\ref{me2}) and (\ref{me4}) coincide and they therefore yield identical physical predictions.  However, in the near-zone $R\ll \lambda_0$ the master equations generally exhibit significant differences. To compare the two sets of predictions we assume a vacuum field $N=0$ and consider the experimental situation in which the system is prepared in the symmetric state $\ket{\epsilon_3}$. This state is a simultaneous eigenstate of the dipole Hamiltonian $\omega_0(\sigma_1^+\sigma_1^-+\sigma_2^+\sigma_2^-)$ appearing in the standard master equation (\ref{me2}), and of the Hamiltonian $H_d$ appearing in our master equation (\ref{me4}). Experimentally, one expects to find that the system initially prepared in the state $\ket{\epsilon_3}$ decays into the stationary state. Theoretically, different stationary states are predicted by the two master equations (\ref{me2}) and (\ref{me4}), and the rates of decay into these respective stationary states are also different. Figs.~\ref{ft} and \ref{f1} compare the symmetric and stationary state populations found using master equations (\ref{me2}) and (\ref{me4}) when the system starts in the symmetric eigenstate $\ket{\epsilon_3}$. For small separations the ground and symmetric state populations obtained from our master equation (\ref{me4}) crossover earlier, which indicates more rapid symmetric state decay than is predicted by Eq.~(\ref{me2}) (see Fig.~\ref{ft}). This gives rise to the different starting values at $R=r_a$ of the curves depicted in Fig.~\ref{f1}. For larger separations the solutions converge and become indistinguishable for all times.

\begin{figure}
\begin{minipage}{\columnwidth}
\begin{center}
\includegraphics{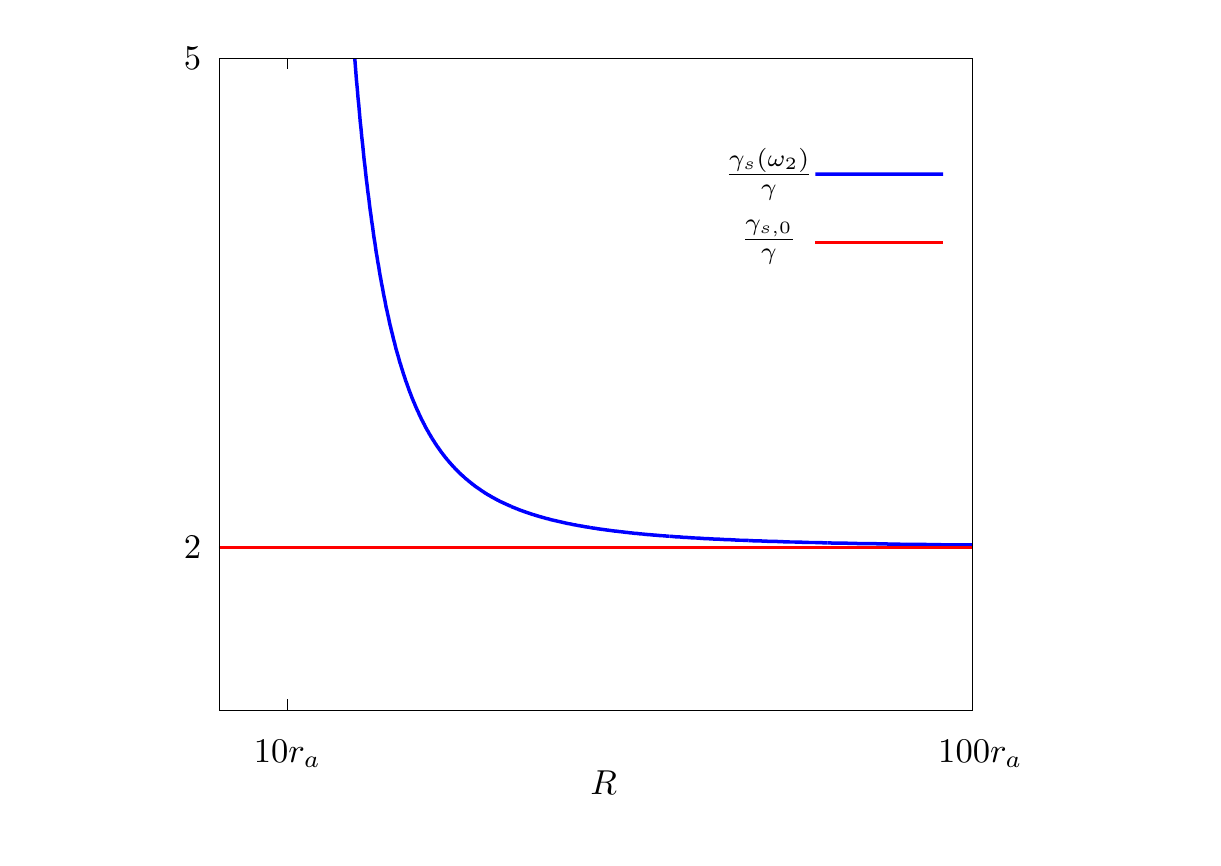}
\vspace*{-0.5cm}
\caption{Comparison of the symmetric state decay rates $\gamma_s$ (from our master equation) and $\gamma_{s,0}$ (from the standard master equation) as functions of separation $R$. All parameters are chosen as in Fig.~\ref{ft}.}\label{fG}
\end{center}
\end{minipage}
\end{figure}

The different behaviour in Figs.~\ref{ft} and \ref{f1} can be understood by looking at a few relevant quantities. The matrix element of the combined dipole moment between ground and symmetric eigenstates is found to be
\begin{align}\label{d31}
{\bf d}_{31} = \bra{\epsilon_3} {\bf d}_1+{\bf d}_2 \ket{\epsilon_1} = 2a{\bf d},
\end{align}
which is different to the usual transition dipole moment $\bra{\epsilon_3} {\bf d}_1+{\bf d}_2 \ket{gg} =\sqrt{2}{\bf d}$. Since $a \to 1/\sqrt{2}$ as $C\to 0$, the dipole moment ${\bf d}_{31}$ reduces to $\sqrt{2}{\bf d}$ when $R\to \infty$. As $R$ decreases, however, ${\bf d}_{13}$ becomes increasingly large compared with $\sqrt{2}{\bf d}$. This is consistent with the more rapid decay observed in Fig.~\ref{ft}. A more complete explanation of this behaviour can be obtained by calculating the rate of decay of the symmetric state into the vacuum, which we denote $\gamma_s$. Using Fermi's golden rule, and the eigenstates given in Eq.~(\ref{estates}), we obtain
\begin{align}\label{ds}
\gamma_s(\omega_2) = 2c^2 [\gamma_{\mu\mu}(\omega_2) + \gamma_{12}(\omega_2)]_{N=0}.
\end{align}
Only when $C\to 0$, such that $\omega_2 =\eta +C \to \omega_0$ and $c\to 1/\sqrt{2}$, does this decay rate reduce to that obtained when using the bare eigenstates $\ket{i,j}, (i,j=e,g)$, which is
\begin{align}\label{gams0}
\gamma_{s,0} = \gamma + \gamma_{12}(\omega_0),
\end{align}
where $\gamma_{12}(\omega_0)$ is given in Eq.~(\ref{coeffs1}). As shown in Fig.~\ref{fG}, for sufficiently small $R$ the decay rate $\gamma_s(\omega_2)$ is significantly larger than $\gamma_{s,0}$.

\begin{figure}
\begin{minipage}{\columnwidth}
\begin{center}
\includegraphics{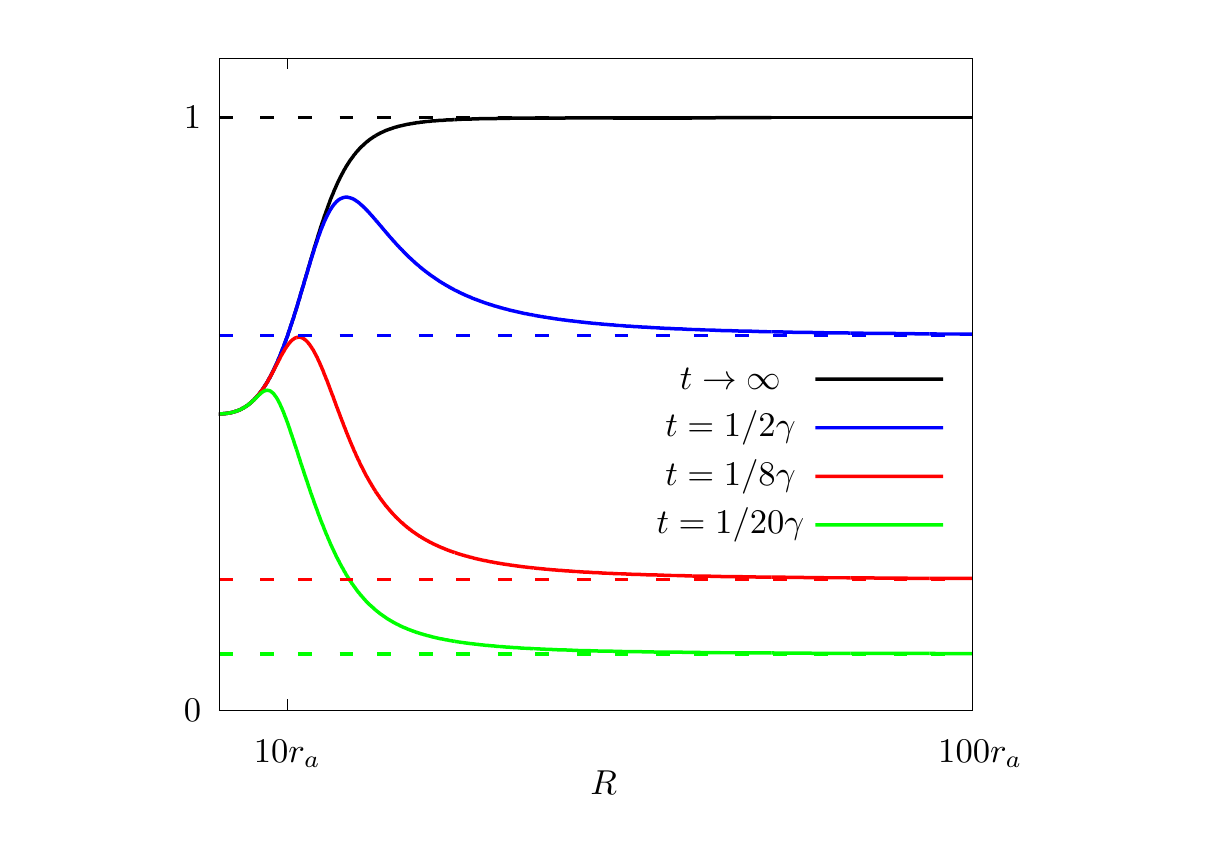}
\vspace*{-0.5cm}
\caption{The population of the state $\ket{g,g}$ found using Eq.~(\ref{me4}) is plotted as a function of separation $R$ for various times. All remaining parameters are as in Fig.~\ref{ft}. The dashed lines give the corresponding populations found using Eq.~(\ref{me2}), which are insensitive to variations in $R$ over the range considered. For large $R$ the two sets of solutions agree. In particular, the steady state population $p_{g,g}(\infty) = |\braket{g,g|\epsilon_1}|^2$ is equal to unity only for sufficiently large $R$.}\label{fgg}
\end{center}
\end{minipage}
\end{figure}

In contrast to the decay behaviour of the symmetric state, the predictions of the master equations (\ref{me2}) and (\ref{me4}) are the same if the system is assumed to be prepared in the anti-symmetric state $\ket{\epsilon_2}$, which like $\ket{\epsilon_3}$ is a simultaneous eigenstate of $\omega_0(\sigma_1^+\sigma_1^-+\sigma_2^+\sigma_2^-)$ and $H_d$. Both master equations predict that the population of the state $\ket{\epsilon_2}$ remains stationary, i.e., that it is a completely dark state. This can be understood by noting that the collective dipole moment associated with the anti-symmetric to stationary state transition vanishes when either stationary state, $\ket{g,g}$ or $\ket{\epsilon_1}$, is used. Finally, the predicted behaviour by our master equation (\ref{me4}) of the standard stationary state $\ket{g,g}$ is illustrated in Fig.~\ref{fgg}. For an initial state $\ket{\epsilon_3}$ the population $p_{gg}(t)$ of the state $\ket{g,g}$ at a given time $t$, is identical to that predicted by Eq.~(\ref{me2}) only for sufficiently large $R$ whereby $\ket{\epsilon_1} \approx \ket{g,g}$.
%%%%%%%%%%%%%%%%%%%%%%%%%%%%%%%%%%%%%%%%%%%%%%%%%%%%%%%%%%%%%%%%%
%
%	F I G U R E S  S T A R T
%
%%%%%%%%%%%%%%%%%%%%%%%%%%%%%%%%%%%%%%%%%%%%%%%%%%%%%%%%%%%%%%%%%%

%%%%%%%%%%%%%%%%%%%%%%%%%%%%%%%%%%%%%%%%%%%%%%%%%%%%%%%%%%%%%%%%%%
%
%	F I G U R E S  E N D
%
%%%%%%%%%%%%%%%%%%%%%%%%%%%%%%%%%%%%%%%%%%%%%%%%%%%%%%%%%%%%%%%%%%

\subsection{Emission spectrum}\label{4}

In this section we apply our master equation (\ref{me4}) to calculate the emission spectrum of the two-dipole system initially prepared in the symmetric state $\ket{\epsilon_3}$. This provides a means by which to test experimentally 
%within an experiment
whether our predictions are closer to measured values than the standard approach. 
%, which predictions are closest to the measured values. 
The spectrum of radiation is defined according to the quantum theory of photodetection by \cite{glauber_quantum_1963,carmichael_statistical_2003}
\begin{align}\label{spec}
s(\omega) = \int_0^\infty dt \int_0^\infty dt' e^{-i\omega(t-t')} \langle {\bf E}_{s,{\rm rad}}^{(-)}(t,{\bf x})\cdot {\bf E}^{(+)}_{s,{\rm rad}}(t',{\bf x})\rangle_0,
\end{align}
where for simplicity we assume that the field is in the vacuum state. Since the master equations (\ref{me2}) and (\ref{me4}) yield different predictions for this experimentally measurable quantity, an experiment could be used to test which master equation is the most accurate.

The detector is located at position ${\bf x}$ with $x\gg R$, so that only the radiative component ${\bf E}_{s, {\rm rad}}$ of the electric source field, which varies as $|{\bf x}-{\bf R}_\mu|^{-1}$, need be used. 
%The radiative part of the field 
This is the only part of the field responsible for irreversibly carrying energy away from the sources. The positive and negative frequency components of the radiation source field are given within both rotating-wave and Markov approximations by
\begin{align}\label{elec}
E_{s,{\rm rad},i}^{(\pm)}(t,{\bf x}) = \sum_{\mu=1}^2 {\omega_0^2 \over 4\pi r_\mu}(\delta_{ij}-{\hat r}_{\mu,i}{\hat r}_{\mu,j})d_j\sigma^{\mp}(t_\mu),
\end{align} 
where ${\bf r}_\mu = {\bf x}-{\bf R}_\mu$, and $t_\mu = t-r_\mu$ is the retarded time associated with the $\mu$'th source. For $x\gg R$ we have to a very good approximation that ${\bf r}_1 = {\bf r}_2 = {\bf r}$, where ${\bf r}$ is the relative vector from ${\bf x}$ to the midpoint of ${\bf R}_1$ and ${\bf R}_2$. Substituting Eq.~(\ref{elec}) into Eq.~(\ref{spec}) within this approximation yields
\begin{align}\label{spec2}
s(\omega) = \mu \omega_0^4 \int_0^\infty dt \int_0^\infty dt' e^{-i\omega(t-t')} \sum_{\mu,\nu=1}^2 \langle \sigma_\mu^+(t)\sigma_\nu^-(t')\rangle,
\end{align}
where
\begin{align}
\mu = \left[{1\over 4\pi r}(\delta_{ij}-{\hat r}_i{\hat r}_j)d_j\right]^2.
\end{align}

To begin with, let us use the standard master equation (\ref{me2}) to find the required two-time correlation function. Assuming that the system is initially prepared in the symmetric state $\ket{\epsilon_3}$, the standard master equation (\ref{me2}) together with the method of calculation given in appendix \ref{ap3} we obtain the two-time correlation function
\begin{align}\label{corr3}
\sum_{\mu,\nu=1}^2 \langle \sigma_\mu^+(t)\sigma_\nu^-(t')\rangle = 2e^{-{\gamma_{s,0}\over 2}(t+t')}e^{i({\tilde \omega}_0+\Delta_{12})(t-t')},
\end{align}
where $\gamma_{s,0}$ and $\Delta_{12}$ are given in Eqs.~(\ref{gams0}) and (\ref{coeffs1}), respectively. By direct integration of Eq.~(\ref{corr3}) one obtains the corresponding Lorentzian spectrum
\begin{align}\label{spec3}
s_0(\omega) = {2\omega_0^4 \mu \over (\gamma_{s,0}/2)^2+(\omega-[{\tilde \omega}_0+\Delta_{12}])^2}.
\end{align}

Let us now turn our attention to the spectrum obtained from our new master equation (\ref{me4}). We have seen that the solutions of Eqs.~(\ref{me4}) and (\ref{me2}) differ only in the near field regime $R\ll\lambda_0$. For sufficiently small $R$ we have that $\omega_0 \sim C$, and the frequency difference $\omega_2-\omega_1=2C \sim 2\omega_0$ is large. In this situation we can perform a full secular approximation within Eq.~(\ref{me4}) to obtain the master equation
\begin{align}\label{me5}
{\dot \rho} = -i[{\tilde H}_d,\rho]+{\mathcal D}(\rho),
\end{align}
with
\begin{align}
{\tilde H}_d = H_d +\hspace*{-3mm}\sum_{\omega =\pm \omega_{1,2}}\sum_{\mu,\nu=1}^2 S_{\mu\nu}(\omega)A_{\mu\omega}^\dagger A_{\nu\omega},
\end{align}
and
\begin{align}
{\mathcal D}(\rho) =\hspace*{-3mm}\sum_{\omega =\pm \omega_{1,2}} \sum_{\mu,\nu=1}^2 \gamma_{\mu\nu}(\omega) \left[A_{\nu\omega}\rho A_{\mu\omega}^\dagger - {1\over 2}\{A_{\mu\omega}^\dagger A_{\nu\omega},\rho\}\right].
\end{align}
Solving this secular master equation allows us to obtain a simple expression for the emission spectrum.

The correlation function in Eq.~(\ref{spec}) defines the radiation intensity when it is evaluated at $t=t'$. Naively calculating the quantity $\sum_{\mu,\nu=1}^2\langle \sigma^+_\mu(t)\sigma^-_\nu(t)\rangle_1$ taken in the stationary state $ \ket{\epsilon_1}$ yields a non-zero stationary intensity, because $\ket{\epsilon_1}$ is a superposition involving both $\ket{g,g}$ and the doubly excited bare state $\ket{e,e}$. A non-zero radiation intensity even in the stationary state is clearly non-physical. However, a more careful analysis recognises that when the radiation source fields are to be used in conjunction with Eq.~(\ref{me4}) the optical approximations used in their derivation should be applied in the interaction picture defined in terms of the dressed Hamiltonian $H_d$ given in Eq.~(\ref{hd}). One then obtains the source field
\begin{align}\label{elec2}
E_{s,{\rm rad},i}^{(+)}&(t,{\bf x}) = \sum_{\mu=1}^2 \sum_{\substack{nm \\n<m}} {\epsilon_{nm}^2 \over 4\pi r_\mu}(\delta_{ij}-{\hat r}_{\mu,i}{\hat r}_{\mu,j})d_j\sigma_{\mu,nm} \theta_{nm}(t_\mu),
\end{align}
where $\sigma_{\mu,nm} = \sigma^+_{\mu,nm}+\sigma^-_{\mu,nm}$ in which $\sigma^\pm_{\mu,nm}$ denotes the $nm$'th matrix element of $\sigma^\pm_\mu$ in the basis $\ket{\epsilon_n}$. The transition frequencies associated with this basis are denoted $\epsilon_{nm} = \epsilon_n -\epsilon_m$, and the raising and lowering operators are denoted $\theta_{nm}=\ket{\epsilon_n}\bra{\epsilon_m},~n\neq m$. The derivation of Eq.~(\ref{elec2}) is given in Appendix \ref{appel}. According to Eq.~(\ref{elec2}) the annihilation (creation) radiation source field is now associated with lowering (raising) operators in the dressed basis $\ket{\epsilon_i}$ rather than in the bare basis $\ket{n,m}, (n,m=e,g)$. Substitution of Eq.~(\ref{elec2}) into Eq.~(\ref{spec}) yields
\begin{align}\label{corr5}
s(\omega) = \mu \int_0^\infty dt \int_0^\infty dt' \sum_{\mu,\nu=1}^2 \sum_{\substack{nm \\n<m}}&\sum_{\substack{pq \\q<p}} \epsilon_{pq}^2 \epsilon_{nm}^2 \sigma_{\mu,pq}\sigma_{\nu,nm}  \langle \theta_{pq}(t)\theta_{nm}(t')\rangle,
\end{align}
where we have again assumed that ${\bf r}_1 ={\bf r}_2 = {\bf r}$. Unlike the correlation function in Eq.~(\ref{corr2}), when $t=t'$ the correlation functions $ \langle \theta_{pq}(t)\theta_{nm}(t')\rangle,~p>q,~m>n$ vanish in the stationary (ground) state $\theta_{11}$. The radiation intensity is therefore seen to vanish in the stationary limit as required physically.

Taken in the symmetric state $\theta_{33}$ the only non-zero two-time correlation function that contributes to Eq.~(\ref{corr5}) is found to be
\begin{align}\label{C33}
C_{33}(t,t')=\langle \theta_{31}(t)\theta_{13}(t')\rangle=e^{-c^2[\gamma_{\mu\mu}(\omega_2)+\gamma_{12}(\omega_2)](t+t')}e^{i{\tilde \omega}_2(t-t')},
\end{align} 
where
\begin{align}\label{C33rates}
{\tilde \omega}_2=\omega_2 + 2\big(&S_{\mu\mu}(-\omega_1)[b^2-a^2]+S_{12}(-\omega_1)[b^2+a^2]\nonumber \\ &+c^2[S_{\mu\mu}(\omega_2)-S_{\mu\mu}(-\omega_2)+S_{12}(\omega_2)-S_{12}(-\omega_2)]\big)
\end{align} 
is the shifted symmetric to ground transition frequency. Integration of Eq.~(\ref{C33}) according to Eq.~(\ref{corr5}) then yields the Lorentzian spectrum
\begin{align}\label{specfin}
s(\omega)={(2a\omega_2^2)^2\mu \over (\gamma_s(\omega_2)/2)^2+(\omega-{\tilde \omega}_2)^2}.
\end{align} 
Full details of the calculation of the spectrum in Eq.~(\ref{specfin}) are given in appendix \ref{appel}. In the limit of large separation $C\to 0$,  which implies that $\omega_2 \to \omega_0$, ${\tilde \omega}_2\to {\tilde \omega}_0+\Delta_{12}$, $\gamma_s(\omega_2)\to \gamma_{s,0}$, and $a\to 1/\sqrt{2}$. As a result $s(\omega)\to s_0(\omega)$ for large $R$ and the predicted spectra coincide. On the other hand, for sufficiently small $R$ the spectrum $s(\omega)$ again offers separation-dependent corrections to the standard result $s_0(\omega)$.

%%%%%%%%%%%%%%%%%%%%%%%%%%%%%%%%%%%%%%%%%%%%%%%%%%%%%%%%%%%%%%%%%%
%
%	F I G U R E S  S T A R T
%
%%%%%%%%%%%%%%%%%%%%%%%%%%%%%%%%%%%%%%%%%%%%%%%%%%%%%%%%%%%%%%%%%%
\begin{figure}
\begin{minipage}{\columnwidth}
\begin{center}
\includegraphics{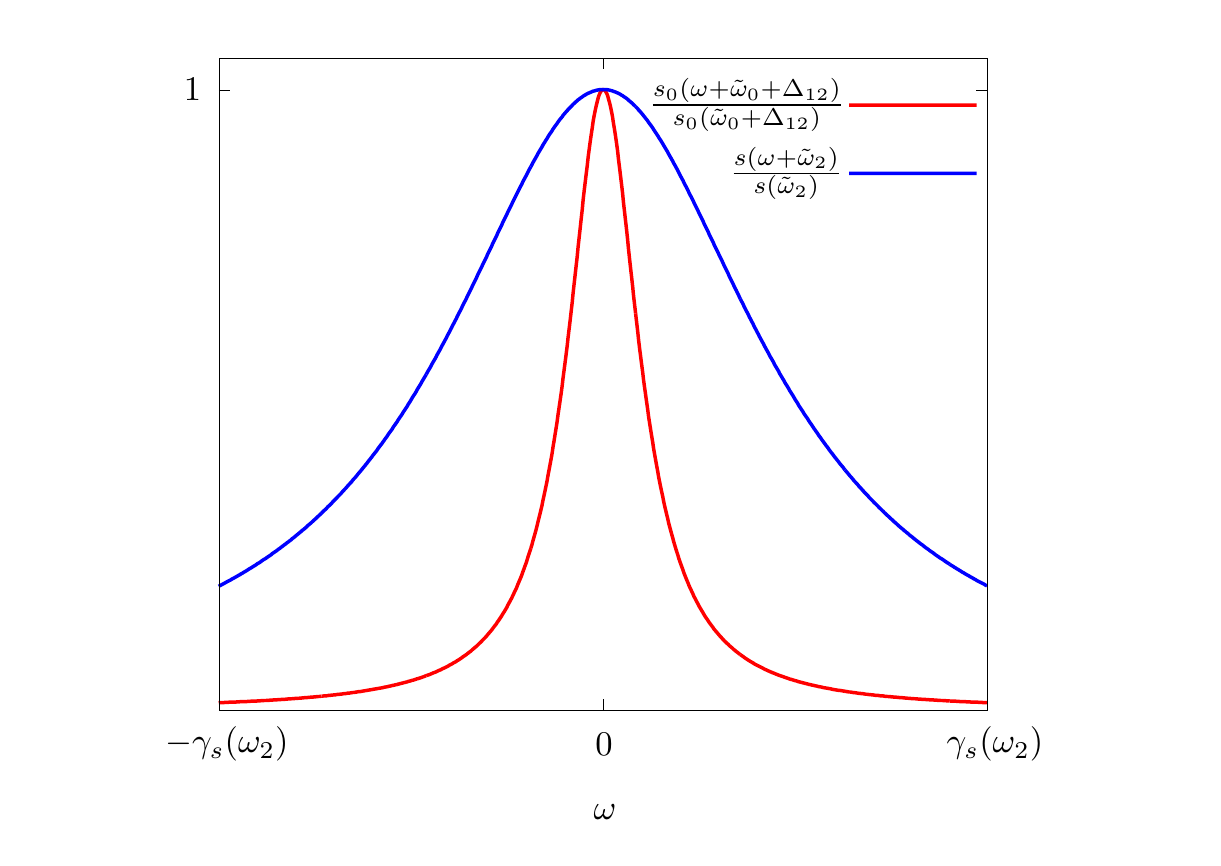} 
\vspace*{-0.5cm}
\caption{The spectra $s(\omega)$ and $s_0(\omega)$ are plotted with $R=10r_a$ and with all remaining parameters as in Fig.~\ref{ft}. For this separation the peak heights and centres are quite different as shown in Figs.~\ref{fh} and \ref{fw}, respectively. Here, for illustrative purposes, the spectra have both been centred at zero and normalised by their respective peak heights.}\label{fsp1}
\end{center}
\end{minipage}
\end{figure}

\begin{figure}
\begin{minipage}{\columnwidth}
\begin{center}
\includegraphics{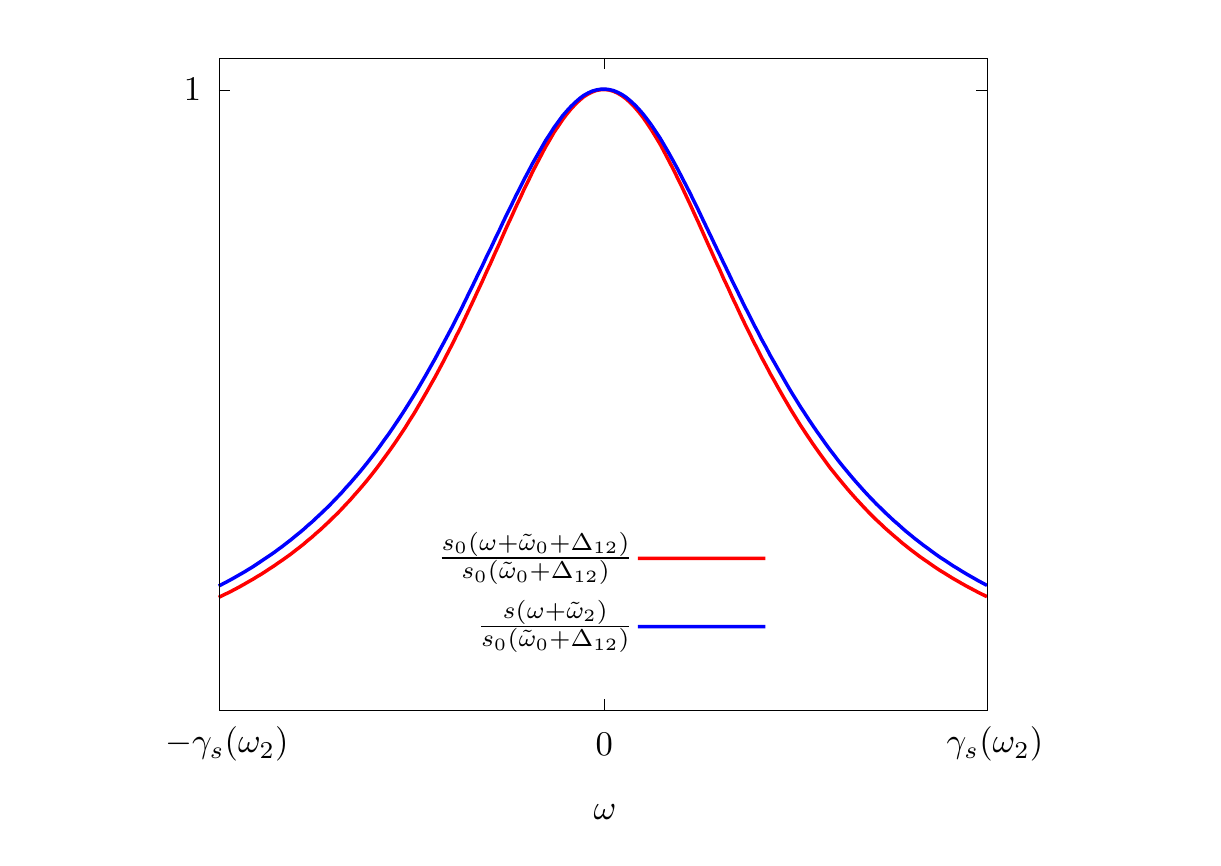} 
\vspace*{-0.5cm}
\caption{The spectra $s(\omega)$ and $s_0(\omega)$ are plotted with $R=50r_a$ and with all remaining parameters as in Fig.~\ref{ft}. For this separation the positions of the peak centres remain quite different on the frequency scale set by the width $\gamma_s(\omega_2)\approx \gamma_{s,0}$. Here, for illustrative purposes, the spectra have both been centred at zero. However, for this value of $R$ the peak heights are effectively the same. Therefore, the spectra have been normailsed by the same peak value $s_0({\tilde \omega}_0+\Delta_{12})$.}\label{fe}
\end{center}
\end{minipage}
\end{figure}

\begin{figure}
\begin{minipage}{\columnwidth}
\begin{center}
\includegraphics{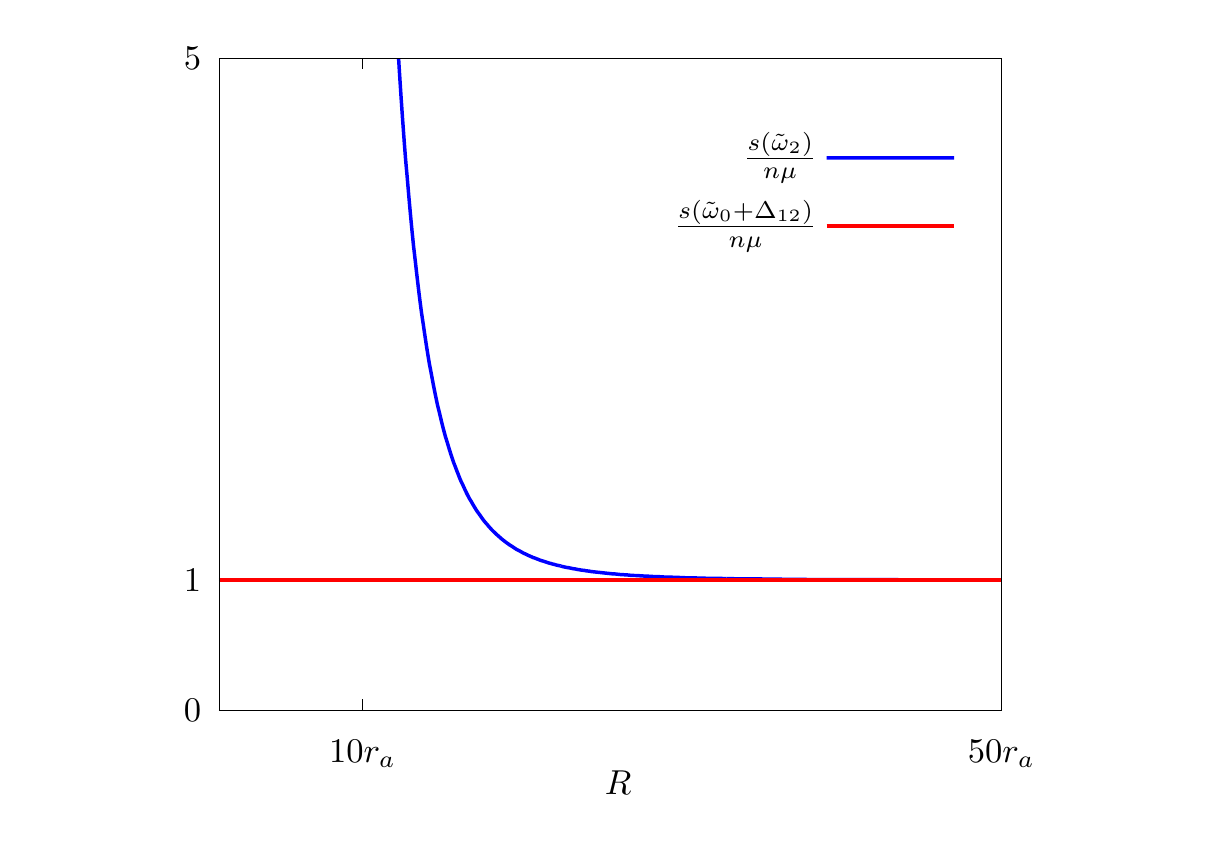}
\vspace*{-0.5cm}
\caption{The relative heights of the peaks in the spectra $s(\omega)$ and $s_0(\omega)$ as a function of the separation $R$. We have chosen a normalisation factor $n=s_0({\tilde \omega_0})|_{R=50 ra}$. We have chosen all remaining parameters as in Fig.~\ref{ft}.}\label{fh}
\end{center}
\end{minipage}
\end{figure}

\begin{figure}
\begin{minipage}{\columnwidth}
\begin{center}
\includegraphics{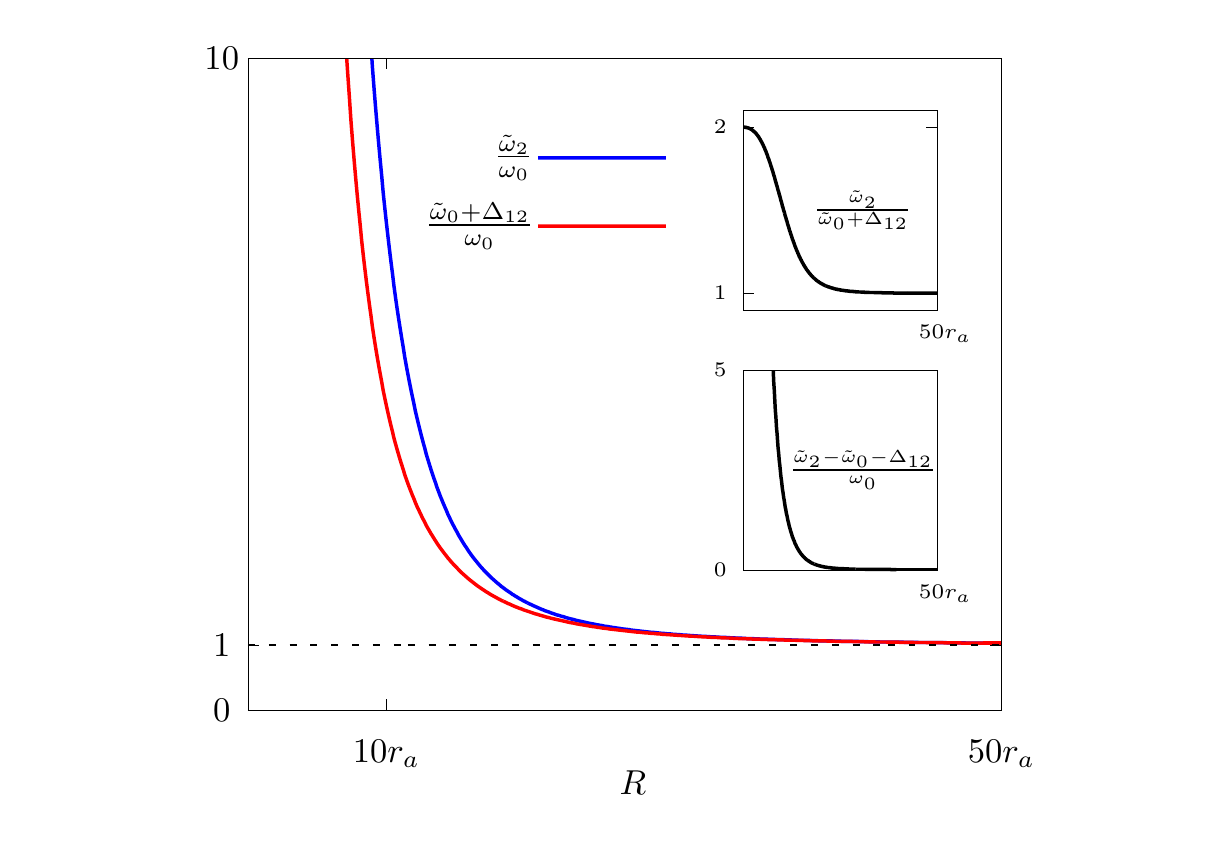}
\vspace*{-0.5cm}
\caption{The positions of the peaks in the spectra $s(\omega)$ and $s_0(\omega)$ as functions of the separation $R$. We have chosen all remaining parameters as in Fig.~\ref{ft}. The upper subplot shows the ratio of the two peak positions over the same range of values of $R$, while the lower subplot shows the difference in peak positions over the same range of values of $R$.}\label{fw}
\end{center}
\end{minipage}
\end{figure}
%%%%%%%%%%%%%%%%%%%%%%%%%%%%%%%%%%%%%%%%%%%%%%%%%%%%%%%%%%%%%%%%%%
%
%	F I G U R E S  E N D
%
%%%%%%%%%%%%%%%%%%%%%%%%%%%%%%%%%%%%%%%%%%%%%%%%%%%%%%%%%%%%%%%%%%

The two spectra $s_0(\omega)$ and $s(\omega)$ are compared in Figs.~\ref{fsp1} and \ref{fe}. As their relative widths are proportional to the rates, they are given in Eqs.~(\ref{gams0}) and (\ref{ds}), respectively. These quantities have been plotted already in Fig.~\ref{fG}. The relative heights of the spectral peaks are $s_0({\tilde \omega}_0+\Delta_{12})/\mu$ and $s({\tilde \omega}_2)/\mu$ respectively, which are plotted in Fig.~\ref{fh}. This figure shows that the peak heights in the spectra begin to diverge as $R$ decreases. At a separation of $15 r_a$, where $r_a= n^2 a_0,~n=50$ is a characteristic Rydberg atomic radius, the peak value of $s(\omega)$ is around two times larger than the peak value of $s_0(\omega)$ for the parameters chosen here. The positions of the peaks are ${\tilde \omega}_0+\Delta_{12}$ and ${\tilde \omega}_2$, respectively, and these are plotted in Fig.~\ref{fw}. The ultra-violet cut-off chosen for the calculation of the single-dipole shift components corresponds to the inverse dipole radius wavelength, namely $2\pi c /r_a$. This value is chosen for consistency with the electric dipole approximation that we have used throughout. For small $R$ the spectrum $s(\omega)$ is blue-shifted relative to $s_0(\omega)$. Fig.~\ref{fw} shows that the ratio of peak positions approaches a constant value around two for very small $R$. These differences could in principle be detected in an experiment. At a separation of $20r_a$, which is roughly $2.5\mu$m, for instance, the difference in shifted frequencies ${\tilde \omega}_2-{\tilde \omega}_0-\Delta_{12}$ is around $1$ Ghz for the parameters chosen in Fig.~\ref{ft}. This is similar in magnitude to the Lamb-shift in atomic Hydrogen.

\section{Conclusions}\label{5}

In this paper we have derived a partially secular master equation valid for arbitrarily separated dipoles within a common radiation field at arbitrary temperature. The equation is intended for the modelling of dipolar systems in which static dipole-dipole interactions are strong compared with the coupling to transverse radiation. This situation can arise in systems of Rydberg atoms and other molecular systems \cite{raimond_spectral_1981,reinhard_level_2007,vogt_electric-field_2007,altiere_dipole-dipole_2011,ravets_coherent_2014,zhelyazkova_probing_2015,bachor_addressing_2016,kumar_collective_2016,dyachkov_dipoledipole_2016,browaeys_experimental_2016,petrosyan_grover_2016,saffman_quantum_2016,dunning_recent_2016,paris-mandoki_tailoring_2016,hettich_nanometer_2002,mlynek_observation_2014,kim_exciton_2016,higgins_superabsorption_2014}.

We have shown that the standard gauge-invariant two-dipole master equation can only be derived in gauges other than the multipolar gauge if the direct inter-dipole Coulomb energy is included within the interaction Hamiltonian rather than the unperturbed part. Our arbitrary gauge approach makes a particular limitation of this method clear. Specifically, the usual approach can only be justified when the direct Coulomb interaction is weak along with the coupling to transverse radiation. In situations in which this is not the case our master equation, which is based on a repartitioning of the Hamiltonian into unperturbed and interaction parts, yields significant corrections to previous results. In addition to corrections to the decay of the excited states of the system, we have found corrections to the natural emission spectrum of the initially excited system. In principle, spectroscopy could be used to determine which predictions are closer to the measured values. A possible extension of our result %consists in including 
would be to include an external driving Hamiltonian that represents coherent irradiation. The techniques employed here could then be used to calculate the fluorescence spectrum of the driven system.

{\em Acknowledgment:} This work was supported by the Engineering and Physical Sciences Research Council. We thank Jake Iles-Smith and Victor Jouffrey for useful discussions.

\section{Appendix}

\subsection{Self-energy contributions and the Gauge-invariance of the single dipole-shift}\label{ap1}

Here we determine the contribution of self-energy terms to dipole level-shifts and demonstrate that the single-dipole transition shift is gauge-invariant. The self-energy term $V^{(2)}$ is given in Eq. (\ref{V2}). The shifts arising from this term are divergent in the mode continuum limit $\omega_k\to \infty$, but this divergence is not unexpected within the non-relativistic dipole approximated treatment. It is typically handled through the introduction of an ultra-violet cut-off. In the treatment of the Lamb-shift in atomic Hydrogen the Coulomb gauge self-energy $V^{(2)}$ with $\alpha_k=0$ is independent of the atomic electron levels and is therefore ignored within the calculation of the measurable shift \cite{craig_molecular_1984}. In the multipolar gauge $V^{(2)}$ represents a polarisation self-energy term and when its contribution is combined with the remaining contribution to the shift coming from the linear part of the multipolar interaction Hamiltonian one obtains the same result as the Coulomb gauge treatment. In all cases mass renormalisation must also be performed to obtain the correct shift.

In the Coulomb gauge $V^{(2)}$ does not contribute to the master equation transition shift of the two-level dipole, which is the difference between excited and ground state shifts. This is independent of whether the two-level approximation has been made. However, even within the Coulomb gauge it is important to note that one must generally account for all self-energy contributions when explicitly verifying that quantities are gauge-invariant. In particular, to verify that the ground and excited level-shifts are separately gauge-invariant, the contributions $\bra{n}^{(2)}V\ket{n},~n=e,g$ must be taken into account.

Using the Hamiltonian in Eq.~(\ref{ham3}), the standard Born-Markov master equation has the form given in Eq.~(\ref{me1}), in which the decay rate $\gamma$ is independent of $\alpha_k$. The transition shift expressed as the difference between excited and ground state shifts as ${\tilde \omega}_0 -\omega_0=\Delta = \delta^{(1)}_e - \delta^{(1)}_g$ where
\begin{align}\label{bs}
\delta^{(1)}_e = \int d^3k &\sum_\lambda {|{\bf e}_\lambda({\bf k})\cdot {\bf d}|^2 \over 2(2\pi)^3} \omega_0\left[{{u_k^+}^2N_k \over \omega_k+\omega_0} -{{u_k^-}^2[1+N_k]\over \omega_k-\omega_0}\right], \nonumber \\
\delta^{(1)}_g = \int d^3k &\sum_\lambda {|{\bf e}_\lambda({\bf k})\cdot {\bf d}|^2 \over 2(2\pi)^3}\omega_0 \left[{{u_k^-}^2N_k \over \omega_k-\omega_0} -{{u_k^+}^2[1+N_k]\over \omega_k+\omega_0}\right],
\end{align}
are $\alpha_k$-dependent. This $\alpha_k$-dependence is due to the lack of any contribution from the self-energy term $V^{(2)}$ in Eq.~(\ref{bs}).

The $\alpha_k$-dependence within the master equation is eliminated when one accounts for the self-energy contributions and the effect of the two-level approximation, recalling that the latter was made after the transformation $R_{\{\alpha_k\}}$ was performed. More specifically it is possible to demonstrate that the single dipole master equation (\ref{BM3}) is $\alpha_k$-independent, and that it coincides with Eq.~(\ref{me1}). First we note that we can continue to express the second line in Eq.~(\ref{BM3}) in terms of the original partition $H=H_0+V$. Thus, provided $H_0$ is kept the same for each choice of the $\alpha_k$ the dissipative part of the master equation is $\alpha_k$-independent.

It remains to show that when one adds the shift contributions $\delta_{e,g}^{(1)}$ coming from the second line in Eq.~(\ref{BM3}) to the corresponding self-energy contribution in Eq.~(\ref{ses}) one obtains gauge-invariant total shifts. To this end let us first consider the Coulomb gauge $\alpha_k=0$. The total excited and ground state shifts are
\begin{align}
\delta_{e,{\rm CG}} = \delta_{e,{\rm CG}}^{(1)}+\delta_{\rm CG}^{(2)}, \qquad
\delta_{g,{\rm CG}} = \delta_{g,{\rm CG}}^{(1)}+\delta_{\rm CG}^{(2)}.
\end{align}
The components $\delta_{e,{\rm CG}}^{(1)}$ are obtained by setting $\alpha_k=0$ in Eq.~(\ref{bs}), while the remaining component
\begin{align}\label{cgs}
\delta_{\rm CG}^{(2)} = {e^2\over 2m}\int d^3 k\sum_\lambda {|{\bf e}_\lambda({\bf k})|^2\over 2(2\pi)^3\omega_k}(1+2N_k)
\end{align}
is the Coulomb gauge self-energy shift due to the ${\bf A}_{\rm T}^2$ part of the Coulomb gauge interaction Hamiltonian. Since this term is independent of the dipole, the shift $\delta_{\rm CG}^{(2)}$ is the same for the ground and excited levels. The single-dipole transition shift $\Delta$ given in Eq.~(\ref{rates1}) in the main text can be expressed in terms of Coulomb gauge shifts as
\begin{align}\label{del23}
\Delta = \delta_{e,{\rm CG}} - \delta_{g,{\rm CG}} =\delta_{e,{\rm CG}}^{(1)} - \delta_{g,{\rm CG}}^{(1)}.
\end{align}
More generally, for arbitrary $\alpha_k$ the total ground and excited state level shifts are denoted $\delta_{e,g}$. In what follows we will show that
\begin{subequations}
\begin{equation}\label{cruc1}
\delta_e - \delta_{\rm CG}^{(2)} = \delta_{e,{\rm CG}}^{(1)} 
\end{equation}
and    
\begin{equation}\label{cruc2}
\delta_g - \delta_{\rm CG}^{(2)} = \delta_{g,{\rm CG}}^{(1)},
\end{equation}
\end{subequations}
from which it follows using Eq.~(\ref{del23}) that $\delta_e-\delta_g=\Delta$ for all choices of $\alpha_k$.

In order to show that Eqs.~(\ref{cruc1}) and (\ref{cruc2}) hold we must carefully account for the two-level approximation, which was performed after the gauge transformation $R_{\{\alpha_k\}}$. Let us consider a general shift of the $m$'th level of the dipole with the form
\begin{align}
{\tilde \omega}_m = \omega_m + \sum_n \omega_{nm}|{\bf v}\cdot {\bf d}_{nm}|^2,
\end{align}
where ${\bf v}$ is arbitrary. If we restrict ourselves to two levels $e$ and $g$, and if $m=e$ in the above, then the sum includes only one other level $n=g$, so we get for the shift
\begin{align}\label{a}
\sum_n \omega_{ne}|{\bf v}\cdot {\bf d}_{ne}|^2 = -\omega_0|{\bf v}\cdot {\bf d}|^2,
\end{align}
where $\omega_0:=\omega_{eg}=-\omega_{ge}$ and ${\bf d}:={\bf d}_{eg}={\bf d}_{ge}^*$. If instead $m=g$ then the shift is
\begin{align}\label{b}
\sum_n \omega_{ng}|{\bf v}\cdot {\bf d}_{ng}|^2= +\omega_0|{\bf v}\cdot {\bf d}|^2.
\end{align}
The shift is clearly different in the $m=e$ and $m=g$ cases when considering a two-level system. However, for an infinite-dimensional dipole the shift is independent of $m$ being given by
\begin{align}\label{c}
\sum_n \omega_{nm}|{\bf v}\cdot {\bf d}_{nm}|^2 = {e^2\over 2m}|{\bf v}|^2,
\end{align}
where we have made use of the identity
\begin{align}\label{id}
\sum_n \omega_{nm}d_{nm}^i d_{mn}^j = i{e^2\over 2m}\bra{m}[p_i,r_j]\ket{m} = \delta_{ij} {e^2\over 2m}.
\end{align}
The difference between the finite and infinite-dimensional cases arises because the proof of Eq.~(\ref{id}) rests directly on the CCR algebra $[r_i,p_j]=i\delta_{ij}$, which can only be supported in infinite-dimensions. When the algebra is truncated to ${su}(2)$, the same shift comes out level-dependent. Since the gauge transformation $R_{\{\alpha_k\}}$ is made on the infinite-dimensional dipole it is necessary to employ Eq.~(\ref{c}) in order to exhibit gauge-invariance of the shifts. Thus, in order to get the correct level-shifts within the two-level approximation, when dealing with the excited level shift $m=e$ we use Eqs.~(\ref{a}) and (\ref{c}), which imply
\begin{align}\label{2id1}
\omega_0|{\bf v}\cdot {\bf d}|^2=- {e^2\over 2m}|{\bf v}|^2,
\end{align}
but when dealing with the ground level shift $m=g$ we use Eqs.~(\ref{b}) and (\ref{c}), which imply
\begin{align}\label{2id2}
\omega_0|{\bf v}\cdot {\bf d}|^2={e^2\over 2m}|{\bf v}|^2.
\end{align}

We now proceed to verify that Eqs.~(\ref{cruc1}) and (\ref{cruc2}) hold. The complete shifts $\delta_{e,g}$ are obtained by taking the shifts in Eq.~(\ref{bs}) and adding their respective self-energy contributions. Subtracting $\delta_{\rm CG}^{(2)}$ in Eq.~(\ref{cgs}) from $\delta_e$ and subsequently using Eq.~(\ref{2id1}), which is appropriate for the excited state shift, we obtain
\begin{align}\label{sh}
\delta_e -\delta_{\rm CG}^{(2)} = \int d^3k &\sum_\lambda {|{\bf e}_\lambda({\bf k})\cdot {\bf d}|^2 \over 2(2\pi)^3}\nonumber \\ &\times \Bigg(\alpha_k^2 -\alpha_k(\alpha_k-2)[1+2N_k]{\omega_0\over\omega_k} + \omega_0\left[{{u_k^+}^2N_k \over \omega_k+\omega_0} -{{u_k^-}^2[1+N_k]\over \omega_k-\omega_0}\right]\Bigg).
\end{align}
Using Eq.~(\ref{u}) we express the bracket within the integrand in this expression in terms of $\alpha_k$.  The part independent of $N_k$ is
\begin{align}\label{j}
\alpha_k^2-\alpha_k(\alpha_k-2){\omega_0\over \omega_k} -{\omega_0 \over \omega_k -\omega_0}\bigg[(1-\alpha_k)^2{\omega_0\over \omega_k} +\alpha_k^2{\omega_k\over \omega_0}+2\alpha_k(1-\alpha_k)\bigg].
\end{align}
In this expression we identify the coefficient of $\alpha_k^2$ as
\begin{align}
1-{\omega_0\over \omega_k-\omega_0}\left({\omega_k\over \omega_0} +{\omega_0\over \omega_k} -2\right)-{\omega_0\over \omega_k}
=1-{\omega_k-\omega_0 \over \omega_k}-{\omega_0\over \omega_k} =0,
\end{align}
and the coefficient of $2\alpha_k$ as
\begin{align}
{\omega_0\over \omega_k}-{\omega_0\over \omega_k-\omega_0}\left(1-{\omega_0\over \omega_k}\right)
={\omega_0\over \omega_k}-{\omega_0\over \omega_k-\omega_0}{\omega_k-\omega_0 \over \omega_k} =0. 
\end{align}
Thus, Eq.~(\ref{j}) is $\alpha_k$-independent. The remaining part is
\begin{align}
{\omega_0^2 \over \omega_k(\omega_0 - \omega_k)}.
\end{align}

The $N_k$-dependent parts of $\delta_e-\delta_{\rm CG}^{(2)}$ can be dealt with in a similar manner. The coefficient of $\alpha_k^2$ in the $N_k$-dependent part of the bracket within the integrand of the expression for $\delta_e-\delta_{\rm CG}^{(2)}$ is
\begin{align}
&-2{\omega_0\over \omega_k}+\left({\omega_0^2\over \omega_k}+\omega_k\right)\left({1\over \omega_0+\omega_k}-{1\over \omega_k-\omega_0}\right)+\omega_0\left({1\over \omega_0+\omega_k}+{1\over \omega_k-\omega_0}\right) \nonumber \\=& ~2{\omega_0\over \omega_k}\left[-1+{1\over \omega_k^2-\omega_0^2}\left(-\omega_0^2-\omega_k^2+2\omega_k^2\right)\right] =0.
\end{align}
Similarly, the coefficient of $\alpha_k$ is
\begin{align}
&4{\omega_0\over \omega_k}+2{\omega_0^2\over \omega_k}\left({1\over \omega_k-\omega_0}-{1\over \omega_k+\omega_0}\right) -2\left({1\over \omega_k-\omega_0}-{1\over \omega_k+\omega_0}\right) \nonumber \\ =& ~4{\omega_0\over \omega_k}\left[1-{1\over (\omega_0+\omega_k)(\omega_k-\omega_0)}(\omega_k^2-\omega_0^2)\right] = 0.
\end{align}
The remaining $N_k$-dependent part is
\begin{align}\label{k}
{\omega_0^2\over \omega_k}\left({1\over \omega_k+\omega_0} + {1\over \omega_0 -\omega_k}\right).
\end{align}
Combining Eqs.~(\ref{j}) and (\ref{k}) we obtain the $\alpha_k$-independent result
\begin{align}\label{z1}
\delta_e  - \delta_{\rm CG}^{(2)}= \int d^3k &\sum_\lambda {|{\bf e}_\lambda({\bf k})\cdot {\bf d}|^2 \over 2(2\pi)^3} {\omega_0^2 \over \omega_k} \left( {[1+N_k] \over \omega_0-\omega_k} + { N_k \over \omega_0+\omega_k}\right) = \delta_{e,{\rm CG}}^{(1)},
\end{align}
which completes the proof of Eq.~(\ref{cruc1}).

The shift appearing on the left-hand-side of Eq.~(\ref{cruc2}) is found using Eq.~(\ref{2id2}) to be
\begin{align}
\delta_g-\delta_{\rm CG}^{(2)} = \int d^3k &\sum_\lambda {|{\bf e}_\lambda({\bf k})\cdot {\bf d}|^2 \over 2(2\pi)^3} \nonumber \\ &\times \Bigg(\alpha_k^2 +\alpha_k(\alpha_k-2)[1+2N_k]{\omega_0\over\omega_k}+ \omega_0 \left[{{u_k^-}^2N_k \over \omega_k-\omega_0} -{{u_k^+}^2[1+N_k]\over \omega_k+\omega_0}\right]\Bigg).
\end{align}
Similar calculations to those above for the excited state yield the final result 
\begin{align}\label{z2}
\delta_g-\delta_{\rm CG}^{(2)} = -\int d^3k \sum_\lambda {|{\bf e}_\lambda({\bf k})\cdot {\bf d}|^2 \over 2(2\pi)^3} {\omega_0^2\over \omega_k}\left( {[1+N_k] \over \omega_0+\omega_k} + {N_k \over \omega_0-\omega_k}\right)=\delta_{g,{\rm CG}}^{(1)}.
\end{align}
This completes the proof that the transition shift $\delta_e-\delta_g$ is $\alpha_k$-independent and that it equals $\Delta$ given in Eq.~(\ref{rates1}).

We remark that the need to account for the self-energy contributions along with the effect of the two-level truncation is a peculiarity of the single-dipole shift term $\Delta$. The same need does not arise in the case of the remaining coefficients $\gamma$, $\gamma_{12}$ and $\Delta_{12}$ in the standard two-dipole master equation (\ref{me2}). These coefficients are immediately seen to coincide with gauge-invariant matrix elements.

\subsection{Calculation of the standard joint shift}\label{ap2}

The joint shift $\Delta_{12}$ resulting from the arbitrary gauge master equation derivation is given by
\begin{align}\label{coeffs2} 
\Delta_{12}=\int {d^3k\over (2\pi)^3} \sum_\lambda |{\bf e}_{{\bf k}\lambda}\cdot {\bf d}|^2e^{i{\bf k}\cdot {\bf R}} \left(\alpha_k^2-1 - {\omega_0\over 2}\left[{{u_k^+}^2\over \omega_k+\omega_0} + {{u_k^-}^2\over \omega_k-\omega_0}\right]\right).
\end{align}
Using Eq.~(\ref{u}) all $\alpha_k$-dependence can be shown to vanish in the same way as with the single-dipole shifts dealt with in Appendix \ref{ap1}. The final result is
\begin{align}
\Delta_{12}=\int {d^3k\over (2\pi)^3} &\sum_\lambda |{\bf e}_\lambda({\bf k})\cdot {\bf d}|^2e^{i{\bf k}\cdot {\bf R}} {\omega_k^2 \over \omega_0^2-\omega_k^2}.
\end{align}
Evaluating the angular integral and polarisation summation yields
\begin{align}\label{delt12}
\Delta_{12}={1\over \pi}\int_0^\infty d\omega_k \, d_i d_j \tau_{ij}(\omega_k,R){\omega_k \over \omega_0^2-\omega_k^2}.
\end{align}
The integral is regularised by introducing a convergence factor $e^{-\epsilon \omega_k}$ under the integral, and finally taking the limit $\epsilon \to 0^+$. We substitute $\tau_{ij}$ given in Eq.~(\ref{tau}) into Eq.~(\ref{delt12}) and evaluate the resulting integrals term by term. The integral arising from the first part of $\tau_{ij}$ is
\begin{align}
\lim_{\epsilon\to 0^+} \int_0^\infty d\omega_k \, \omega_k^3 e^{-\epsilon \omega_k} {\sin \omega_k R \over \omega_0^2-\omega_k^2} = {1\over 2i}\lim_{\epsilon\to 0^+} \int_{-\infty}^\infty d\omega_k \, \omega_k^3 e^{-\epsilon \omega_k} {e^{i\omega_k R} \over \omega_0^2-\omega_k^2}.
\end{align}
We now make the substitution $z= \omega_k R$, and make a suitable choice of contour $C$ such that by the residue theorem we obtain
\begin{align}
{1\over 2i R^2}\lim_{\epsilon\to 0^+} \int_C dz \, { z^3 e^{i z - \epsilon z/R} \over (\omega_0 R)^2-z^2}= -{\pi \over 2}\omega_0^2 \cos \omega_0 R.
\end{align}
Thus, the part of the shift $\Delta_{12}$ arising from the first part ($R^{-1}$ component) of $\tau_{ij}$ is
\begin{align}
-{\omega_0^2\over 4\pi R}(\delta_{ij} -{\hat R}_i {\hat R}_j)d_i d_j \cos\omega_0 R,
\end{align}
which we recognise as the $R^{-1}$ component of $\Delta_{12}$ in Eq.~(\ref{coeffs1}). The remaining parts of Eq.~(\ref{delt12}) can be evaluated in a similar way, which yields the final result given in Eq.~(\ref{coeffs1}).

\subsection{Method of calculation of the spectrum}\label{ap3}

We denote the dynamical map governing evolution of the reduced density matrix by $F(t,t')$, which is such that $F(t,t')\rho(t')=\rho(t)$. A general two-time correlation function for arbitrary system observables $O$ and $O'$ can be written \cite{breuer_theory_2002}
\begin{align}\label{corr0}
\langle O(t)O'(t')\rangle = {\rm tr}(OF(t,t')O'F(t')\rho).
\end{align}
We define the super-operator $\Lambda$ by ${\dot \rho}(t) = \Lambda \rho(t)$ using the master equation [Eq.~(\ref{me2}) or Eq.~(\ref{me4})]. Since $\Lambda$ is time-independent, from the initial condition $F(0,0)\equiv I$ we obtain the general solution $F(t,t') = e^{\Lambda (t-t')}$. For convenience we write $F(t,0)=F(t)$, so that $F(t,t')=F(t-t')$.

In order to calculate the two-time correlation functions we first find a concrete representation of the maps $\Lambda$ and $F(t)$. For this purpose we introduce a basis of operators denoted $\{x_i : i=1,...,16\}$, which is closed under Hermitian conjugation. The trace defines an inner-product $\langle O,O'\rangle = {\rm tr}(O^\dagger O')$ with respect to which the basis $x_i$ is assumed to be orthonormal. We identify two resolutions of unity as $\sum_i {\rm tr} (x_i^\dagger \cdot ) x_i = I = \sum_i {\rm tr} (x_i \cdot ) x_i^\dagger$, which imply that any operator $O$ can be expressed as $O= \sum_i {\rm tr}(x^\dagger_i O)x_i =\sum_i {\rm tr}(x_i O)x^\dagger_i$. Expressing both sides of the equation ${\dot F}(t)=\Lambda F(t)$ in the basis $x_i$ yields the relation
\begin{align}\label{F}
{\dot F}_{jk}(t) = \sum_l \Lambda_{jl} F_{lk}(t),
\end{align}
where
\begin{align}\label{Fjk}
F_{jk}(t)={\rm tr}[x^\dagger_j F(t)x_k],\qquad \Lambda_{jl} = {\rm tr}[x_j^\dagger \Lambda x_l].
\end{align}
Eq.~(\ref{F}) can be written in the matrix form ${\dot {\bf F}} ={\bf \Lambda}{\bf F}(t)$ whose solution is expressible in the matrix exponential form ${\bf F}(t) = e^{{\bf \Lambda}t}$. A general two-time correlation function of system operators can then be expressed using Eq.~(\ref{corr0}) as
\begin{align}\label{corr2}
\langle O(t)O'(t')\rangle =\sum_{ijkl} {\rm tr}(Ox_i)F_{ij}(t-t'){\rm tr}(x_j^\dagger O'x_k)F_{kl}(t'){\rm tr}(x_l^\dagger \rho) 
= {\bf O}^{\rm T}{\bf F}(t-t'){\bf O}'{\bf F}(t'){\bm \rho},
\end{align}
where $O_i = {\rm tr}(Ox_i)$, $\rho_i = {\rm tr}(x_i^\dagger \rho)$ and $O'_{ij} = {\rm tr}(x_i^\dagger O'x_j)$. Choosing the basis $\{x_i\}$ to be the operators obtained by taking the outer products of the bare states $\ket{n,m},~(n,m=e,g)$, the above machinery can be used to obtain the correlation function (\ref{corr3}). 

\subsection{Derivation of spectrum associated with the new master equation}\label{appel}

The mode expansion for the transverse field canonical momentum ${\bf \Pi}_{\rm T}$ is
\begin{align}
{\bf \Pi}_{\rm T}(t,{\bf x}) = -i\sum_{{\bf k}\lambda} \sqrt{\omega_k\over 2 L^3}{\bf e}_{{\bf k}\lambda}a_{{\bf k}\lambda}(t) e^{i{\bf k}\cdot {\bf x}}+{\rm H.c.} 
\end{align}
This operator represents a different physical observable for each choice of $\alpha_k$, because it does not commute with the generalised gauge transformation $R_{\{\alpha_k\}}$. Similarly the photonic operators $a_{{\bf k}\lambda}$ are implicitly different for each choice of $\alpha_k$. In the multipolar gauge the field canonical momentum coincides with the total electric field away from the sources; ${\bf \Pi}_{\rm T}({\bf x}) = -{\bf E}({\bf x}),~{\bf x}\neq {\bf R}_\mu$. The positive frequency (annihilation) and negative frequency (creation) components of the electric field are therefore defined for ${\bf x}\neq {\bf R}_\mu$ by
\begin{align}\label{epm}
{\bf E}^{(+)}(t,{\bf x}) = i \sum_{{\bf k}\lambda} \sqrt{\omega_k\over 2L^3} {\bf e}_{{\bf k}\lambda} a_{{\bf k}\lambda}(t) e^{i{\bf k}\cdot {\bf x}},\qquad{\bf E}^{(-)}(t,{\bf x})= {\bf E}^{(+)}(t,{\bf x})^\dagger,
\end{align}
where $a_{{\bf k}\lambda}$ is the photon annihilation operator within the multipolar gauge. For a system of two dipoles the integrated Heisenberg equation for the multipolar photon annihilation operator yields the source component
\begin{align}\label{as}
a_{{\bf k}\lambda,s}(t) =\sqrt{\omega_k\over 2 L^3}\sum_{\mu=1}^2  e^{-i{\bf k}\cdot {\bf R}_\mu} \int_0^t dt' e^{-i\omega_k(t-t')}{\bf e}_{{\bf k}\lambda}\cdot{\bf d}_\mu(t').
\end{align}
Since the dipole moment operators ${\bf d}_\mu$ commute with the transformation $R_{\{\alpha_k\}}$ they represent the same physical observable for each choice of $\alpha_k$. This implies that Eq.~(\ref{as}) can be expressed in terms of Coulomb gauge raising and lowering operators $\sigma^\pm_\mu$ in the two-level approximation, despite the implicit difference between these operators and their counterparts defined within the multipolar gauge. We subsequently express the Coulomb gauge operators $\sigma^\pm_\mu$ in the dressed basis $\ket{\epsilon_n}$ to obtain
\begin{align}
a_{{\bf k}\lambda,s}(t) = \sqrt{\omega_k\over 2 L^3}{\bf e}_{{\bf k}\lambda}\cdot{\bf d}\sum_{\mu=1}^2 e^{-i{\bf k}\cdot {\bf R}_\mu} \int_0^t dt' e^{-i\omega_k(t-t')}\sum_{nm} \sigma_{\mu,nm}\theta_{nm}(t'),
\end{align}
where $\sigma_{\mu,nm}=\sigma^+_{\mu,nm}+\sigma^-_{\mu,nm}$, $\epsilon_{nm}=\epsilon_n-\epsilon_m$, and $\theta_{nm}=\ket{\epsilon_n}\bra{\epsilon_m}$. We now perform a rotating-wave approximation, which eliminates terms that are rapidly oscillating within the interaction picture defined by the dressed Hamiltonian $H_d$ given in Eq.~(\ref{hd}). Substitution of the resulting expression into Eq.~(\ref{epm}) yields in the mode continuum limit
\begin{align}
{\bf E}^{(+)}_s&(t,{\bf x}) \nonumber \\ &= i\int d^3k \sum_\lambda {\omega_k \over 2(2\pi)^3}{\bf e}_\lambda({\bf k})[{\bf e}_\lambda({\bf k})\cdot {\bf d}] \sum_{\mu=1}^2 \sum_{\substack{nm \\n<m}} e^{i{\bf k}\cdot {\bf r}_\mu}\int_0^t dt' e^{-i\omega_k(t-t')}e^{i\epsilon_{nm}t'} \sigma_{\mu,nm}{\tilde \theta}_{nm}(t'),
\end{align}
where ${\tilde \theta}_{nm}(t')$ denotes the operator $\theta_{nm}(t')$ transformed into the interaction picture with respect to $H_d$, and ${\bf r}_\mu = {\bf x}-{\bf R}_\mu$. Performing the angular integration and polarisation summation, and retaining only the radiative component yields
\begin{align}
E^{(+)}_{s,{\rm rad},i}&(t,{\bf x}) \nonumber \\ &= {i\over 4\pi^2} \sum_{\mu=1}^2 \sum_{\substack{nm \\n<m}}(\delta_{ij}-{\hat r}_{\mu,i}{\hat r}_{\mu, j})d_j  \int_0^\infty d\omega_k \int_0^t dt'  \omega_k^2 {\sin(\omega_k r_\mu) \over r_\mu}e^{-i\omega_k(t-t')}e^{i\epsilon_{nm}t'} \sigma_{\mu,nm}{\tilde \theta}_{nm}(t').
\end{align}
Finally, using the Markov approximation
\begin{align}\label{ma}
&\int_0^\infty d\omega_k \, f(\omega_k) e^{i(\omega_k+\epsilon_{nm}) t'} \left[e^{-i\omega_k (t-r_\mu)}- e^{-i\omega_k (t+r_\mu)}\right] \nonumber \\ &\approx  f(\epsilon_{mn}) \int_{-\infty}^\infty d\omega_k \, e^{i(\omega_k+\epsilon_{nm})t'}\left[e^{-i\omega_k (t-r_\mu)}- e^{-i\omega_k (t+r_\mu)}\right] \nonumber \\ &= 2\pi f(\epsilon_{mn})e^{i\epsilon_{nm}t'} [\delta(t'-(t-r_\mu))-\delta(t'-(t+r_\mu))],
\end{align}
valid for a suitably behaved function $f$, we obtain the final result Eq.~(\ref{elec2}) given in the main text. 

To calculate the spectrum according to our master equation (\ref{me4}) we choose the basis of operators $\{x_i\}$ used within the general method laid out in appendix \ref{ap3} as that obtained by taking the outer-products of the basis states $\ket{\epsilon_n}$ given in Eq.~(\ref{estates}). Using Eq.~(\ref{corr2}) we define the array of correlation functions
\begin{align}\label{corrarr}
C_{nmp}(t,t') = \langle x_n^\dagger(t)x_m(t')\rangle_{x_p} = ({\bf F}(t-t'){\bf X}_m {\bf F}(t') )_{np},
\end{align}
where $p$ is restricted to values such that $x_p$ is diagonal, and where the matrix ${\bf X}_m$ has elements $({\bf X}_m)_{jk} = {\rm tr}(x_j^\dagger x_m x_k)$. Taken in the symmetric state $\theta_{33}$ the correlations appearing in Eq.~(\ref{corr5}) are all elements of the array $C_{nm}(t,t')$, which is given by Eq.~(\ref{corrarr}) with $x_p=\theta_{33}$. We choose a labelling whereby the $x_i$ are given by
\begin{align}
&x_i = \ket{\epsilon_1}\bra{\epsilon_i},~~~i=1,...,4, \nonumber \\
&x_i = \ket{\epsilon_2}\bra{\epsilon_{i-4}},~~~i=5,...,8, \nonumber \\
&x_i = \ket{\epsilon_3}\bra{\epsilon_{i-8}},~~~i=9,...,12, \nonumber \\
&x_i = \ket{\epsilon_4}\bra{\epsilon_{i-12}},~~~i=13,...,16.
\end{align}
In this case the only non-zero off-diagonal element of $C_{nm}(t,t')$ is $C_{1,11}(t,t')$ where $x_1=\theta_{11}$ and $x_{11}=\theta_{33}$. Furthermore the diagonal elements $C_{nn}(t,t')$ are zero unless $n$ is odd. It follows that the only non-vanishing correlations in Eq.~(\ref{corr5}) are $C_{33}(t,t')$ and $C_{77}(t,t')$. Moreover, since $\sum_{\mu,\nu=1}^2 \sigma_{\mu,32}\sigma_{\nu,23} =0$ only the term involving $C_{33}(t,t')$ contributes. This term describes correlations associated with the symmetric to ground state transition and is given by Eq.~(\ref{C33}) in the main text. Integration of this correlation function according to Eq.~(\ref{spec}) then yields the spectrum in Eq.~(\ref{specfin}).
\\ \\
{\bf References}
\\
\bibliography{two_dipoles.bib}

\providecommand{\newblock}{}
\begin{thebibliography}{10}
\expandafter\ifx\csname url\endcsname\relax
  \def\url#1{{\tt #1}}\fi
\expandafter\ifx\csname urlprefix\endcsname\relax\def\urlprefix{URL }\fi
\providecommand{\eprint}[2][]{\url{#2}}

\bibitem{london_zur_1930}
London F 1930 {\em Zeitschrift für Physik\/} {\bf 63} 245--279 ISSN 0044-3328
  \urlprefix\url{https://link.springer.com/article/10.1007/BF01421741}

\bibitem{forster_zwischenmolekulare_1948}
Förster T 1948 {\em Annalen der Physik\/} {\bf 437} 55--75 ISSN 1521-3889
  \urlprefix\url{http://onlinelibrary.wiley.com/doi/10.1002/andp.19484370105/abstract}

\bibitem{casimir_influence_1948}
Casimir H~B~G and Polder D 1948 {\em Physical Review\/} {\bf 73} 360--372
  \urlprefix\url{https://link.aps.org/doi/10.1103/PhysRev.73.360}

\bibitem{agarwal_quantum_2012}
Agarwal G~S 2012 {\em Quantum {Optics}\/} (Cambridge, UK: Cambridge University
  Press) ISBN 978-1-107-00640-9

\bibitem{freedhoff_collective_1979}
Freedhoff H~S 1979 {\em Physical Review A\/} {\bf 19} 1132--1139

\bibitem{kilin_cooperative_1980}
Kilin S~J 1980 {\em Journal of Physics B: Atomic and Molecular Physics\/} {\bf
  13} 2653 ISSN 0022-3700
  \urlprefix\url{http://stacks.iop.org/0022-3700/13/i=13/a=023}

\bibitem{griffin_two-atom_1982}
Griffin R~D 1982 {\em Physical Review A\/} {\bf 25} 1528--1534

\bibitem{ficek_two-atom_1990}
Ficek Z and Sanders B~C 1990 {\em Quantum Optics: Journal of the European
  Optical Society Part B\/} {\bf 2} 269 ISSN 0954-8998
  \urlprefix\url{http://stacks.iop.org/0954-8998/2/i=4/a=001}

\bibitem{santos_master_2014}
Santos J~P and Semião F~L 2014 {\em Physical Review A\/} {\bf 89} 022128
  \urlprefix\url{https://link.aps.org/doi/10.1103/PhysRevA.89.022128}

\bibitem{raimond_spectral_1981}
Raimond J~M, Vitrant G and Haroche S 1981 {\em Journal of Physics B: Atomic and
  Molecular Physics\/} {\bf 14} L655 ISSN 0022-3700
  \urlprefix\url{http://stacks.iop.org/0022-3700/14/i=21/a=003}

\bibitem{reinhard_level_2007}
Reinhard A, Liebisch T~C, Knuffman B and Raithel G 2007 {\em Physical Review
  A\/} {\bf 75} 032712
  \urlprefix\url{https://link.aps.org/doi/10.1103/PhysRevA.75.032712}

\bibitem{vogt_electric-field_2007}
Vogt T 2007 {\em Physical Review Letters\/} {\bf 99}

\bibitem{altiere_dipole-dipole_2011}
Altiere E, Fahey D~P, Noel M~W, Smith R~J and Carroll T~J 2011 {\em Physical
  Review A\/} {\bf 84} 053431
  \urlprefix\url{https://link.aps.org/doi/10.1103/PhysRevA.84.053431}

\bibitem{ravets_coherent_2014}
Ravets S, Labuhn H, Barredo D, Béguin L, Lahaye T and Browaeys A 2014 {\em
  Nature Physics\/} {\bf 10} 914--917 ISSN 1745-2473
  \urlprefix\url{file:///Users/macbook/Library/Application%20Support/Zotero/Profiles/uuo5w5h1.default/zotero/storage/8J29HEE3/nphys3119.html}

\bibitem{zhelyazkova_probing_2015}
Zhelyazkova V and Hogan S~D 2015 {\em Physical Review A\/} {\bf 92} 011402
  \urlprefix\url{https://link.aps.org/doi/10.1103/PhysRevA.92.011402}

\bibitem{bachor_addressing_2016}
Bachor P, Feldker T, Walz J and Schmidt-Kaler F 2016 {\em Journal of Physics B:
  Atomic, Molecular and Optical Physics\/} {\bf 49} 154004 ISSN 0953-4075
  \urlprefix\url{http://stacks.iop.org/0953-4075/49/i=15/a=154004}

\bibitem{kumar_collective_2016}
Kumar S, Sheng J, Sedlacek J~A, Fan H and Shaffer J~P 2016 {\em Journal of
  Physics B: Atomic, Molecular and Optical Physics\/} {\bf 49} 064014 ISSN
  0953-4075 \urlprefix\url{http://stacks.iop.org/0953-4075/49/i=6/a=064014}

\bibitem{dyachkov_dipoledipole_2016}
D'yachkov L~G, Zelener B~V, Klyarfeld A~B and Bronin S~Y 2016 {\em Journal of
  Physics: Conference Series\/} {\bf 774} 012162 ISSN 1742-6596
  \urlprefix\url{http://stacks.iop.org/1742-6596/774/i=1/a=012162}

\bibitem{browaeys_experimental_2016}
Browaeys A, Barredo D and Lahaye T 2016 {\em Journal of Physics B: Atomic,
  Molecular and Optical Physics\/} {\bf 49} 152001 ISSN 0953-4075
  \urlprefix\url{http://stacks.iop.org/0953-4075/49/i=15/a=152001}

\bibitem{petrosyan_grover_2016}
Petrosyan D, Saffman M and Mølmer K 2016 {\em Journal of Physics B: Atomic,
  Molecular and Optical Physics\/} {\bf 49} 094004 ISSN 0953-4075
  \urlprefix\url{http://stacks.iop.org/0953-4075/49/i=9/a=094004}

\bibitem{saffman_quantum_2016}
Saffman M 2016 {\em Journal of Physics B: Atomic, Molecular and Optical
  Physics\/} {\bf 49} 202001 ISSN 0953-4075
  \urlprefix\url{http://stacks.iop.org/0953-4075/49/i=20/a=202001}

\bibitem{dunning_recent_2016}
Dunning F~B, Killian T~C, Yoshida S and Burgdörfer J 2016 {\em Journal of
  Physics B: Atomic, Molecular and Optical Physics\/} {\bf 49} 112003 ISSN
  0953-4075 \urlprefix\url{http://stacks.iop.org/0953-4075/49/i=11/a=112003}

\bibitem{paris-mandoki_tailoring_2016}
Paris-Mandoki A, Gorniaczyk H, Tresp C, Mirgorodskiy I and Hofferberth S 2016
  {\em Journal of Physics B: Atomic, Molecular and Optical Physics\/} {\bf 49}
  164001 ISSN 0953-4075
  \urlprefix\url{http://stacks.iop.org/0953-4075/49/i=16/a=164001}

\bibitem{ishizaki_quantum_2010}
Ishizaki A, Calhoun T~R, Schlau-Cohen G~S and Fleming G~R 2010 {\em Physical
  Chemistry Chemical Physics\/} {\bf 12} 7319--7337 ISSN 1463-9084
  \urlprefix\url{http://pubs.rsc.org/en/content/articlelanding/2010/cp/c003389h}

\bibitem{cohen-tannoudji_photons_1989}
Cohen-Tannoudji C, Dupont-Roc J and Grynberg G 1989 {\em Photons and {Atoms}:
  {Introduction} to {Quantum} {Electrodynamics}\/} A {Wiley}-{Interscience}
  publication (Wiley) ISBN 978-0-471-84526-3
  \urlprefix\url{http://books.google.co.uk/books?id=orbvAAAAMAAJ}

\bibitem{stokes_extending_2012}
Stokes A, Kurcz A, Spiller T~P and Beige A 2012 {\em Physical Review A\/} {\bf
  85} 053805 \urlprefix\url{http://link.aps.org/doi/10.1103/PhysRevA.85.053805}

\bibitem{milonni_natural_1989}
Milonni P~W, Cook R~J and Ackerhalt J~R 1989 {\em Physical Review A\/} {\bf 40}
  3764--3768 \urlprefix\url{http://link.aps.org/doi/10.1103/PhysRevA.40.3764}

\bibitem{woolley_gauge_1998}
Woolley R~G 1998 {\em Molecular Physics\/} {\bf 94} 409--416 ISSN 0026-8976
  \urlprefix\url{http://www.tandfonline.com/doi/abs/10.1080/002689798167917}

\bibitem{power_time_1999}
Power E~A and Thirunamachandran T 1999 {\em Physical Review A\/} {\bf 60}
  4936--4942 \urlprefix\url{http://link.aps.org/doi/10.1103/PhysRevA.60.4936}

\bibitem{power_time_1999-1}
Power E~A and Thirunamachandran T 1999 {\em Physical Review A\/} {\bf 60}
  4927--4935 \urlprefix\url{http://link.aps.org/doi/10.1103/PhysRevA.60.4927}

\bibitem{stokes_gauge_2013}
Stokes A 2013 {\em Journal of Physics B: Atomic, Molecular and Optical
  Physics\/} {\bf 46} 145505 ISSN 0953-4075, 1361-6455
  \urlprefix\url{http://iopscience.iop.org/0953-4075/46/14/145505/pdf/0953-4075_46_14_145505.pdf}

\bibitem{woolley_charged_1999}
Woolley R~G 1999 {\em International Journal of Quantum Chemistry\/} {\bf 74}
  531--545 ISSN 1097-461X
  \urlprefix\url{http://onlinelibrary.wiley.com/doi/10.1002/(SICI)1097-461X(1999)74:5<531::AID-QUA9>3.0.CO;2-H/abstract}

\bibitem{stokes_noncovariant_2012}
Stokes A 2012 {\em Physical Review A\/} {\bf 86} 012511
  \urlprefix\url{http://link.aps.org/doi/10.1103/PhysRevA.86.012511}

\bibitem{drummond_unifying_1987}
Drummond P~D 1987 {\em Physical Review A\/} {\bf 35} 4253--4262
  \urlprefix\url{http://link.aps.org/doi/10.1103/PhysRevA.35.4253}

\bibitem{craig_molecular_1984}
Craig D~P and Thirunamachandran T 1984 {\em Molecular quantum electrodynamics:
  an introduction to radiation-molecule interactions\/} (Academic Press) ISBN
  978-0-12-195080-4

\bibitem{breuer_theory_2002}
Breuer H~P, Breuer H~P and Petruccione F 2002 {\em The {Theory} of {Open}
  {Quantum} {Systems}\/} (Oxford ; New York: OUP Oxford) ISBN 978-0-19-852063-4

\bibitem{jaksch_fast_2000}
Jaksch D, Cirac J~I, Zoller P, Rolston S~L, Côté R and Lukin M~D 2000 {\em
  Physical Review Letters\/} {\bf 85} 2208--2211
  \urlprefix\url{https://link.aps.org/doi/10.1103/PhysRevLett.85.2208}

\bibitem{westermann_dynamics_2006}
Westermann S, Amthor T, Oliveira A~L~d, Deiglmayr J, Reetz-Lamour M and
  Weidemüller M 2006 {\em The European Physical Journal D - Atomic, Molecular,
  Optical and Plasma Physics\/} {\bf 40} 37--43 ISSN 1434-6060, 1434-6079
  \urlprefix\url{https://link.springer.com/article/10.1140/epjd/e2006-00130-3}

\bibitem{glauber_quantum_1963}
Glauber R~J 1963 {\em Physical Review\/} {\bf 130} 2529--2539
  \urlprefix\url{http://link.aps.org/doi/10.1103/PhysRev.130.2529}

\bibitem{carmichael_statistical_2003}
Carmichael H~J 2003 {\em Statistical {Methods} in {Quantum} {Optics} 1:
  {Master} {Equations} and {Fokker}-{Planck} {Equations}\/} (New York:
  Springer) ISBN 978-3-540-54882-9

\bibitem{hettich_nanometer_2002}
Hettich C, Schmitt C, Zitzmann J, Kühn S, Gerhardt I and Sandoghdar V 2002
  {\em Science\/} {\bf 298} 385--389 ISSN 0036-8075, 1095-9203
  \urlprefix\url{http://science.sciencemag.org/content/298/5592/385}

\bibitem{mlynek_observation_2014}
Mlynek J~A, Abdumalikov A~A, Eichler C and Wallraff A 2014 {\em Nature
  Communications\/} {\bf 5} ncomms6186 ISSN 2041-1723
  \urlprefix\url{https://www.nature.com/articles/ncomms6186}

\bibitem{kim_exciton_2016}
Kim H, Kim I, Kyhm K, Taylor R~A, Kim J~S, Song J~D, Je K~C and Dang L~S 2016
  {\em Nano Letters\/} {\bf 16} 7755--7760 ISSN 1530-6984
  \urlprefix\url{http://dx.doi.org/10.1021/acs.nanolett.6b03868}

\bibitem{higgins_superabsorption_2014}
Higgins K~D~B, Benjamin S~C, Stace T~M, Milburn G~J, Lovett B~W and Gauger E~M
  2014 {\em Nature Communications\/} {\bf 5} ncomms5705 ISSN 2041-1723
  \urlprefix\url{https://www.nature.com/articles/ncomms5705}

\end{thebibliography}

\end{document}